\RequirePackage{rotating}
\documentclass[fleqn,usenatbib]{mnras}

\usepackage[T1]{fontenc}
\usepackage{ae,aecompl}

\usepackage{newtxtext}
\usepackage[varvw]{newtxmath}

\usepackage{graphicx}	
\usepackage{amsmath}	
\usepackage[british]{babel}
\usepackage[]{units}
\usepackage{booktabs}
\usepackage{xcolor}
\usepackage{microtype}
\hypersetup{pdfauthor={Errani et al.},
            pdftitle={},
            pdfkeywords={},
            bookmarksnumbered=true}

\newcommand{\rmax}{r_\mathrm{mx}}	%
\newcommand{\vmax}{V_\mathrm{mx}}	%
\newcommand{\Mmax}{M_\mathrm{mx}}	%
\newcommand{\tmax}{T_\mathrm{mx}}	%
\newcommand{\kms}{\mathrm{km\,s^{-1}}}		%
\newcommand{\Rh}{R_\mathrm{h}}	%
\newcommand{\Rc}{R_\mathrm{c}}	%
\newcommand{\rh}{r_\mathrm{h}}	%
\newcommand{\diff}{\mathrm{d}}
\newcommand{\Msol}{\mathrm{M_{\odot}}}

\newcommand{\Vc}{V_\mathrm{c}}
\newcommand{\Vh}{V_\mathrm{h}}
\newcommand{\kpc}{\mathrm{kpc}}
\newcommand{\Gyrs}{\mathrm{Gyrs}}
\newcommand{\rperi}{r_\mathrm{peri}}
\newcommand{\tperi}{T_\mathrm{peri}}
\newcommand{\Torb}{T_\mathrm{orb}}
\newcommand{\rapo}{r_\mathrm{apo}}
\newcommand{\maxzero}{_\mathrm{mx0}}
\newcommand{\rt}{{\,:\,}}
\newcommand{\E}{{\mathcal{E}}}

\newcommand{\Rhzero}{{R_\mathrm{h0}}}
\newcommand{\sigmalos}{{\sigma_\mathrm{los}}}

\title[Tidally limited satellite galaxies]{Structure and kinematics of tidally limited satellite galaxies in LCDM}

\author[Errani et al.]{Rapha\"el Errani${}^{1,2,}$\thanks{errani@unistra.fr}, Julio F. Navarro${}^2$, Rodrigo Ibata${}^1$, Jorge Pe\~narrubia${}^3$
\\
$^1$ Universit\'e de Strasbourg, CNRS, Observatoire astronomique de Strasbourg, UMR 7550, F-67000 Strasbourg, France\\
$^2$ Department of Physics and Astronomy, University of Victoria, Victoria, BC V8P 5C2, Canada \\
$^3$ Institute for Astronomy, University of Edinburgh, Royal Observatory, Blackford Hill, Edinburgh EH9 3HJ, UK
}

\date{Accepted 2022 February 17;  in original form 2021 November 10}

\pubyear{2022}
\volume{}

\begin{document}

\label{firstpage}
\pagerange{\pageref{firstpage}--\pageref{lastpage}} \pubyear{2022}
\maketitle

\begin{abstract}
We use $N$-body simulations to model the tidal evolution of dark
matter-dominated dwarf spheroidal galaxies embedded in cuspy
Navarro-Frenk-White subhalos. Tides gradually peel off stars and
dark matter from a subhalo, trimming it down according to their
initial binding energy. This process strips preferentially particles
with long orbital times, and comes to an end when the remaining
bound particles have crossing times shorter than a fraction of the
orbital time at pericentre. The properties of the final stellar
remnant thus depend on the energy distribution of stars in the
progenitor subhalo, which in turn depends on the initial density
profile and radial segregation of the initial stellar component.
The stellar component may be completely dispersed if its
energy distribution does not extend all the way to the subhalo
potential minimum, although a bound dark remnant may remain.  These
results imply that ``tidally-limited'' galaxies, defined as systems
whose stellar components have undergone substantial tidal mass loss,
neither converge to a unique structure nor follow a single tidal
track.  On the other hand, tidally
limited dwarfs do have characteristic sizes and velocity dispersions
that trace directly the characteristic radius ($\rmax$) and circular
velocity ($\vmax$) of the subhalo remnant. This result places strong
upper limits on the size of satellites whose unusually low velocity
dispersions are often ascribed to tidal effects. In particular, the 
large size of kinematically-cold ``feeble giant'' satellites like 
Crater 2 or Antlia 2 cannot be explained as due to tidal effects alone 
in the Lambda Cold Dark Matter scenario.
\end{abstract}





\section{Introduction}
\label{SecIntro}

The Lambda Cold Dark Matter  (LCDM) scenario makes well-defined and falsifiable predictions for the radial density profile of dark matter halos, as well as for their abundance as a function of virial\footnote{We shall define ``virial'' quantities as those measured within spheres of mean density equal to $200\times$ the critical density for closure, $\rho_\mathrm{crit}=3H_0^2/8\pi G$, where $H_0 = 67\, \mathrm{km}\, \mathrm{s}^{-1}\, \mathrm{Mpc}^{-1}$ is the value of Hubble's constant \citep{Planck2020}. Virial quantities are designated with a ``200'' subscript.} mass. These predictions are particularly relevant for the study of dwarf spheroidal (dSph) galaxies, dark matter-dominated systems whose stellar components act as simple kinematic tracers of the structure of their surrounding dark halos \citep[see; e.g.,][]{Mateo1993,Walker2007}.

These studies may be used to test the expected density profiles of cold dark matter halos, which are well approximated by the Navarro-Frenk-White formula \citep[hereafter, NFW;][]{Navarro1996a,Navarro1997}. This issue has been debated for decades, albeit with mixed results, with some authors arguing that the structure of dwarf galaxy halos is consistent with the ``cuspy'' NFW shape and others claiming that the data suggest density profiles with a sizable constant-density ``core'' \citep[for reviews see, e.g.,][]{Gilmore2007,Bullock2017_Review}.

The interpretation of these studies is further complicated by the fact that the assembly of the baryonic component of a galaxy may induce changes in the dark matter density profile, perhaps even erasing the expected cusp and imprinting a core \citep{Navarro1996b,Read2005,Governato2010,Pontzen2012}. However, such baryon-induced effects should be minimal in very faint dark matter-dominated galaxies, simply because the total fraction of mass in baryonic form is too small to be able to affect gravitationally the dark matter component \citep[see; e.g.,][]{Penarrubia2012,DiCintio2014,Benitez-Llambay2019}.

Dwarf galaxies may also be used to probe another robust LCDM prediction, concerning the mass function of low-mass halos. In LCDM these halos are so numerous, and their mass function so steep, that accommodating the comparatively scarcer number of dwarfs and their much shallower stellar mass function requires that galaxy formation efficiency should decline steadily with decreasing halo mass, effectively restricting dwarf galaxy formation to halos in a narrow range of virial mass \citep{Guo2010,Ferrero2012,Sales2017, Bullock2017_Review}.

A firm lower limit to that narrow range ($M_{200}\sim 10^9\, M_\odot$)  is suggested by the minimum virial temperature needed to allow hydrogen to cool efficiently, after accounting for the effects of an ionizing UV background \citep[the ``hydrogen cooling limit'', HCL;][]{Efstathiou1992,Quinn1996,Gnedin2000,Okamoto2009,Benitez-Llambay2020}. In quantitative terms, this implies that essentially all dwarfs with stellar mass, $M_\star<10^7\, M_\odot$, no matter how faint they may be, should form in halos with characteristic circular velocity\footnote{It is customary to use the maximum circular velocity of a halo, $\vmax$, as a proxy for virial mass, $M_{200}$. These two measures are strongly correlated in LCDM via the mass-concentration relation \citep[see; e.g.,][]{Ludlow2016}. The radius at which the circular velocity peaks is usually denoted $\rmax$. This characteristic radius and circular velocity fully specify the structure of an NFW halo.} somewhere between $20\,\kms$ and $40\,\kms$ \citep{Fattahi2018}. 

Together with the NFW profile mentioned above, the HCL minimum mass sets a velocity dispersion ``floor'' for dark matter-dominated dwarfs of given radius, since more massive NFW halos generally have, at all radii, larger circular velocities than less massive ones. This velocity ``floor'' is, to first order, simply a fraction of the circular velocity (at that radius) of a halo at the HCL boundary.  For example, for an HCL halo with $\vmax \approx 20\,\kms$ (reached at a radius $\rmax = 3.4\,\kpc$), the circular velocity at $300\,\mathrm{pc}$ is $\approx 11\,\kms$. This implies that a dwarf with 3D half-light radius, $r_\mathrm{h}\sim 300$ pc should have a line-of-sight (i.e., 1D) velocity dispersion, $\sigma_{\rm los}$, in excess of $11/\sqrt{3}\approx 6\,\kms$. The same argument results in $\sigma_{\rm los} \gtrsim 9\,\kms$ for $r_\mathrm{h}\sim 1\,\kpc$, and $\sigma_{\rm los} \gtrsim4\,\kms$ for $r_\mathrm{h}\sim 100\,\mathrm{pc}$. (These numbers assume a halo of average concentration at $z=0$.)

There are a number of dwarfs in the Local Group that appear to violate these limits \citep[see, e.g., the compilation maintained by][]{McConnachie2012}, a result which, taken at face value, would call for a review of some of the basic assumptions on which the above predictions are based. However, much of the wealth of available kinematic data on dwarfs concerns {\it satellite} galaxies in the Local Group, mainly those orbiting the Milky Way (MW) and the Andromeda (M31) galaxies \citep[for a review, see][]{Simon2019}.

The predictions described above do not apply to satellites, as tides arising from the MW and M31 may strip a significant fraction of the total dark matter content of a dwarf while leaving the stellar component relatively undisturbed \citep{Penarrubia2008,EPT15,Sanders2018}. Translating the LCDM predictions described above to the realm of satellite galaxies thus requires a good understanding of how the dark and stellar components of dSphs evolve as a result of tidal effects.

This is an issue that has been addressed in earlier work, resulting in  a number of conclusions and suggestions about how to interpret kinematic and photometric data on Local Group satellites in the context of LCDM. Two highlights of that work include the suggestion that (i) the stellar mass, size, and velocity dispersion of dSphs evolve along well-specified ``tidal tracks'' which depend only on the total amount of mass lost from within the stellar half-light radius \citep{Penarrubia2008}, and that (ii) the stellar components of ``tidally limited'' satellites, defined as those whose stellar mass has been substantially reduced by tides, approach a ``Plummer-like'' density profile shape roughly independent of the initial distribution of stars \citep{Penarrubia2009}.

These conclusions, however, were based on simulations exploring a rather limited set of initial conditions, in terms of the assumed initial dSph structure, and also of the assumed radial segregation between stars and dark matter. Considering a broader range of possibilities for either may result in revised predictions that could impact, in particular, the properties of ``tidally limited'' dwarfs. This is important, especially in light of the recent discovery of a population of dwarfs with unusually large sizes and low velocity dispersions, well below the limits mentioned earlier \citep[see, e.g., the Crater 2 and Antlia 2 dSphs;][]{Torrealba2016,CaldwellWalker2017,Torrealba2019}.

Although it is tempting to associate such systems with tidally-limited dwarfs \citep{Frings2017,Sanders2018,Amorisco2019}, it is important to realize that the extreme estimated tidal losses put them in a regime hitherto unexplored in previous work. For example, \citet{Fattahi2018} argue that Crater 2 may be the result of a system that has lost more than $99\%$ of its stars and has seen its velocity dispersion reduced by a factor of $\sim 5$. As discussed by \citet{EN21}, exploring this regime with N-body simulations requires resolving the progenitor subhalo with over $10^7$ particles, well beyond what has been achieved with cosmological hydrodynamical simulations.

As a result of these limitations, many questions regarding the structure and survival of tidally-limited galaxies remain unanswered. Recent work, for example, has argued convincingly that NFW subhalos are only very rarely fully disrupted by tides, almost always leaving behind self-bound dark remnants that are missed in cosmological simulations of limited resolution \citep{Kazantzidis2004, Goerdt2007, Penarrubia2010, vdb2018}. How do these results affect the structure and survival of the stellar components of such systems? Do the stellar components of dSphs also survive to some extent (giving rise to ``micro-galaxies'', as argued in \citealt{EP20}), and, if so, what are their properties? Should we expect them all to have similar density profiles? Do they evolve along well-defined ``tidal tracks'' in luminosity, size, and velocity dispersion? Shoud we expect a large population of tidally-limited dwarfs of extremely low surface brightness awaiting discovery?

We explore some of these issues here using a large suite of $N$-body simulations designed to study the tidal evolution of NFW halos in the gravitational potential of a much more massive system, with particular emphasis on the regime of extreme tidal mass loss. These simulations extend our earlier work on the subject \citep{EN21}, where we focussed on the survival and structure of the dark component of the tidal remnant. Our emphasis here is on the evolution of a putative stellar component, assumed to be gravitationally unimportant compared to the dark matter. In the interest of simplicity, we only consider spherical, isotropic models in this contribution, but our approach should be relatively straightforward to extend to include modifications to these assumptions.

\section{Numerical methods}
\label{sec:nummethods}
We use the set of high-resolution $N$-body simulations of the tidal evolution of NFW dark matter subhalos introduced in \citet{EN21} and briefly summarized in this section.

\subsection{Subhalo model} 
\label{sec:subhalo-model}
We model subhalos hosting dSph galaxies as $N$-body realisations of NFW \citep{Navarro1996a,Navarro1997} density profiles,
\begin{equation}
\label{eq:NFW}
 \rho_\mathrm{NFW}(r) = {\rho_\mathrm{s} \over \left(r/r_\mathrm{s} \right) \left(1+ r/r_\mathrm{s} \right)^{2}},
\end{equation}
where $\rho_\mathrm{s}$ and $r_\mathrm{s}$ are a scale density and scale radius, respectively. The NFW profile has a circular velocity curve $V_\mathrm{c}(r) = \sqrt{G M(<r)/r}$ that reaches a peak velocity of $\vmax \approx 1.65\,r_\mathrm{s}\, \sqrt{G\rho_\mathrm{s}}$ at a radius $\rmax \approx 2.16\,r_\mathrm{s}$. The NFW density profile is fully specified by the two parameters, which may be taken to be $\{\rmax,\vmax\}$. We shall hereafter refer to them as the ``characteristic radius'' and ``characteristic velocity'' of the subhalo, respectively. For future reference, we also introduce at this point the subhalo characteristic mass, $\Mmax = \rmax \vmax^2 /G$, and characteristic timescale, $\tmax = 2 \pi \, \rmax / \vmax$. 

As the density profile of Eq.~\ref{eq:NFW} leads to a diverging cumulative mass for $r\rightarrow\infty$, we truncate the profile exponentially at $10\,r_\mathrm{s}$. 
We generate $10^7$-particle equilibrium $N$-body realisations of NFW halos with isotropic velocity dispersion, drawn from a distribution function computed numerically using Eddington inversion, following the implementation\footnote{\label{github}The code to generate $N$-body models and corresponding stellar tagging probabilities is available online at \url{https://github.com/rerrani/nbopy}.} described by \citet{EP20}.

\subsection{Host halo}
We study the tidal evolution of the $N$-body subhalo in an analytical, static, isothermal spherical host potential,
\begin{equation}
  \Phi_\mathrm{host} = V_0^2 \ln(r/r_0)~,
\end{equation}
where $V_0 = 220\,\kms$ denotes the (constant) circular velocity, and $r_0$ is an arbitrary reference radius. The circular velocity curve is flat, with a value $V_0$ chosen to match roughly that of the Milky Way \citep[see, e.g., ][]{Eilers2019}. These parameters correspond to a virial mass at redshift $z=0$ of $M_{200} = 3.7 \times 10^{12}\,\Msol$, and a virial radius of $r_{200} = 325\,\kpc$, respectively.

\subsection{Orbits}

We focus on the evolution of subhalos on eccentric orbits with peri-to-apocentre ratio of $1:5$. This value is consistent with the average peri-to-apocentre ratio of Milky Way satellite galaxies \citep[][table 1]{Li2021}, and close to the typical peri-to-apocentre ratio for dark matter substructures determined by \citet{vdb1999} for isotropic distributions. As discussed by \citet{EN21}, the orbital eccentricity affects mainly the rate of tidal stripping, but not the remnant structure.

If we assume an apocentric distance of $\rapo = 200\,\kpc$, a $1:5$ orbit implies a pericentric distance of $\rperi = 40\,\kpc$, and an orbital period of $\Torb = 2.5\,\Gyrs$.  Note that gravitational effects are scale-free, so all dimensional quantities are just quoted for illustration, and may be rescaled as needed.
  
We shall only explore the properties of subhalos at their orbital apocentres, where they spend most of their orbital time, and when they are closest to equilibrium. Large deviations from equilibrium are expected near pericentre, as discussed by \citet{Penarrubia2009}, but we defer their study to future contributions. 

We show in Appendix~\ref{appendix:eccentricity} that all conclusions derived for the $1:5$ orbit also hold for subhalos on circular orbits, and for more radial orbits with peri-to-apocentre ratios of $1:10$ and $1:20$.

\subsection{Subhalo tidal evolution}
\label{SecTidalEvol}
All subhalos have initial characteristic velocities substantially smaller than that of the host halo, i.e., $V\maxzero/V_0 \lesssim 0.02$. This choice allows us to neglect the effects of dynamical friction, and should be appropriate for studying the evolution of the subhalos of faint and ultrafaint satellites of  galaxies like the MW or M31.

The simulations explore a wide range of initial characteristic radii, $r\maxzero$, chosen so that the ratio of initial subhalo crossing time, $T\maxzero$, to the circular orbital time at at the pericentre of the orbit, $\tperi=2\pi\, \rperi/V_0$, lies in the range $2/3 < T\maxzero / \tperi < 2$.

As discussed by \citet{EN21}, this choice implies heavy mass losses due to tidal stripping. After a few orbits, tidal effects gradually slow down and eventually become negligible once the remnant approaches a characteristic timescale, $\tmax \approx \tperi/4$. The structure of the remnant approaches asymptotically that of an exponentially truncated cusp,
\begin{equation}
\label{eq:TruncCusp}
 \rho_\mathrm{asy}(r) = {\rho_\mathrm{cut}\,  e^{-r/r_\mathrm{cut}} \over \left( r/r_\mathrm{cut} \right)}, 
\end{equation}
with $\rho_\mathrm{cut} $ and $r_\mathrm{cut}$ given by the total mass of the bound remnant, $M_\mathrm{asy}=4\pi\, \rho_\mathrm{cut}\, r_\mathrm{cut}^3$.

As a subhalo is tidally stripped, its characteristic radius and velocity decrease along well-defined tidal tracks \citep{Hayashi2003, Penarrubia2008, EPT15, Green2019}, which may be parameterized by 
\begin{equation}
\vmax/V\maxzero=2^\alpha  (\rmax/r\maxzero)^\beta\, [1+ (\rmax/r\maxzero)^2]^{-\alpha}, 
\label{eq:EN21_tracks}
\end{equation}
with $\alpha=0.4$, $\beta=0.65$ \citep{EN21}.

The structural properties of bound subhalo remnants are determined by first (i) locating the subhalo centre through the shrinking spheres method \citep{Power2003}; then, (ii) assuming spherical symmetry, computing potential and kinetic energies for  $N$-body particles in a frame of reference co-moving with the subhalo; and, then, (iii) removing unbound particles in this co-moving frame. We iterate these steps until convergence.

\subsection{Simulation code}
\label{sec:simcode}
The $N$-body models are evolved using the particle-mesh code \textsc{superbox} \citep{Fellhauer2000}. This code employs a high- and a medium-resolution grid of $128^3$ cells each, co-moving with the subhalo centre of density, with cell size $\Delta x \approx \rmax / 128$ and $\approx 10\, \rmax / 128$, respectively. A third, low-resolution grid is fixed in space with a cell size of $\approx 500\,\kpc/128$. The time integration is performed using a leapfrog-scheme with a time step of $ \Delta t = \min(\tmax, \tperi)/400$. This choice of time step ensures that, for NFW density profiles, at radii corresponding to the cell size of the highest-resolving grid, circular orbits are still resolved by (at least) $\approx 16$ steps. 

The convergence tests of \citet[][appendix A]{EN21} for the set of simulations used in the present work suggest that the structural parameters $\{\rmax,\vmax\}$ of the $N$-body models may be considered unaffected by resolution artefacts as long as $\rmax > 8 \Delta x$ (grid resolution being the main limitation for this set of simulations), which translates to a minimum remnant bound mass fraction of $\Mmax / M\maxzero \approx 1/300$.

\subsection{Stellar tracers}
\label{sec:methods_Stars}
We embed stars as massless tracers using the distribution function-based approach of \citet{Bullock2005} in the implementation\footnotemark[3] of \citet{EP20}. For spherical, isotropic models (the only kind we consider here), the distribution function depends on the phase-space coordinates only through the energy $E=v^2/2 + \Phi(r)$. We associate to each $N$-body dark matter particle in the initial conditions a probability
\begin{equation}
\label{eq:stellartag}
 \mathcal{P}_\star(E) = \left(\diff N_\star/\diff E \right) / \left(\diff N_\mathrm{DM}/\diff E \right)
\end{equation}
of being drawn from a (stellar) energy distribution $\diff N_\star/\diff E $.
Using these tagging probabilities, $\mathcal{P}_\star(E)$, as appropriately normalized weights, the properties of different stellar tracer components can be inferred directly from the $N$-body particles. As stars are modelled as massless tracers of the underlying dark matter potential in this work, we make no distinction between stellar mass $M_\star$ and stellar luminosity~$L$.
Subhalo masses quoted in the text hence refer to the dark matter mass of the bound remnants.

\begin{figure}
 \centering
  \includegraphics[width=8.5cm]{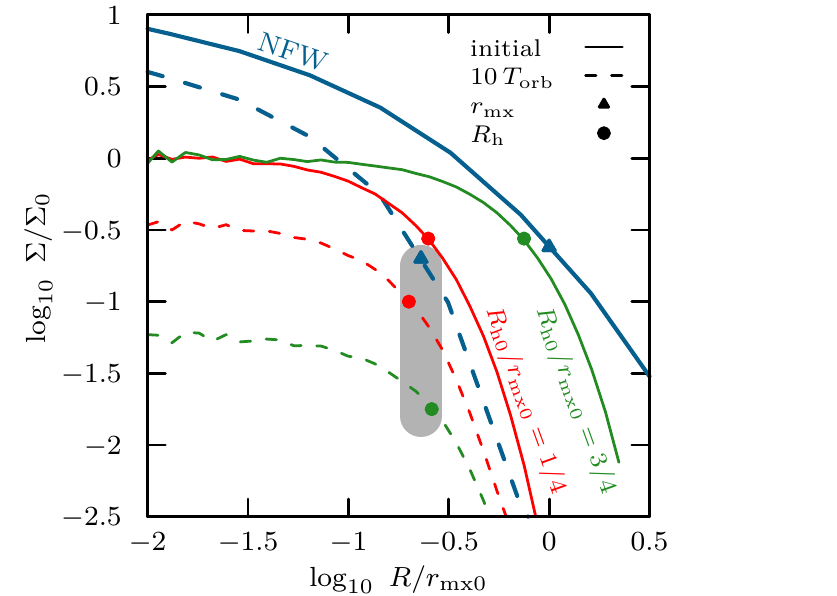}
\caption{Surface density profiles of stars (models \emph{s-i} and \emph{s-ii}) and dark matter (model \emph{h-i}) corresponding to the illustrative case shown in  Fig.~\ref{fig:overview_brightness_evol}. The relative density normalization between stars and dark matter is arbitrary. Initial profiles are shown with solid lines; those after 10 orbital periods with dashed lines. Projected half-light radii, $\Rh$, as well as the characteristic radius of the underlying dark matter component are highlighted using filled circles and filled triangles, respectively. Two different stellar tracers are shown, one with $R_\mathrm{h0} / r\maxzero = 3/4$ (model \emph{s-i}), and a more segregated one with $R_\mathrm{h0} / r\maxzero = 1/4$ (model \emph{s-ii}). After ten orbital periods, both stellar profiles have similar half-light radii, $\Rh  \approx \rmax$, as highlighted by the grey shaded band.  }
\label{fig:initial_final_profs}
\end{figure}

\begin{figure*}
 \centering
  \includegraphics[width=\textwidth]{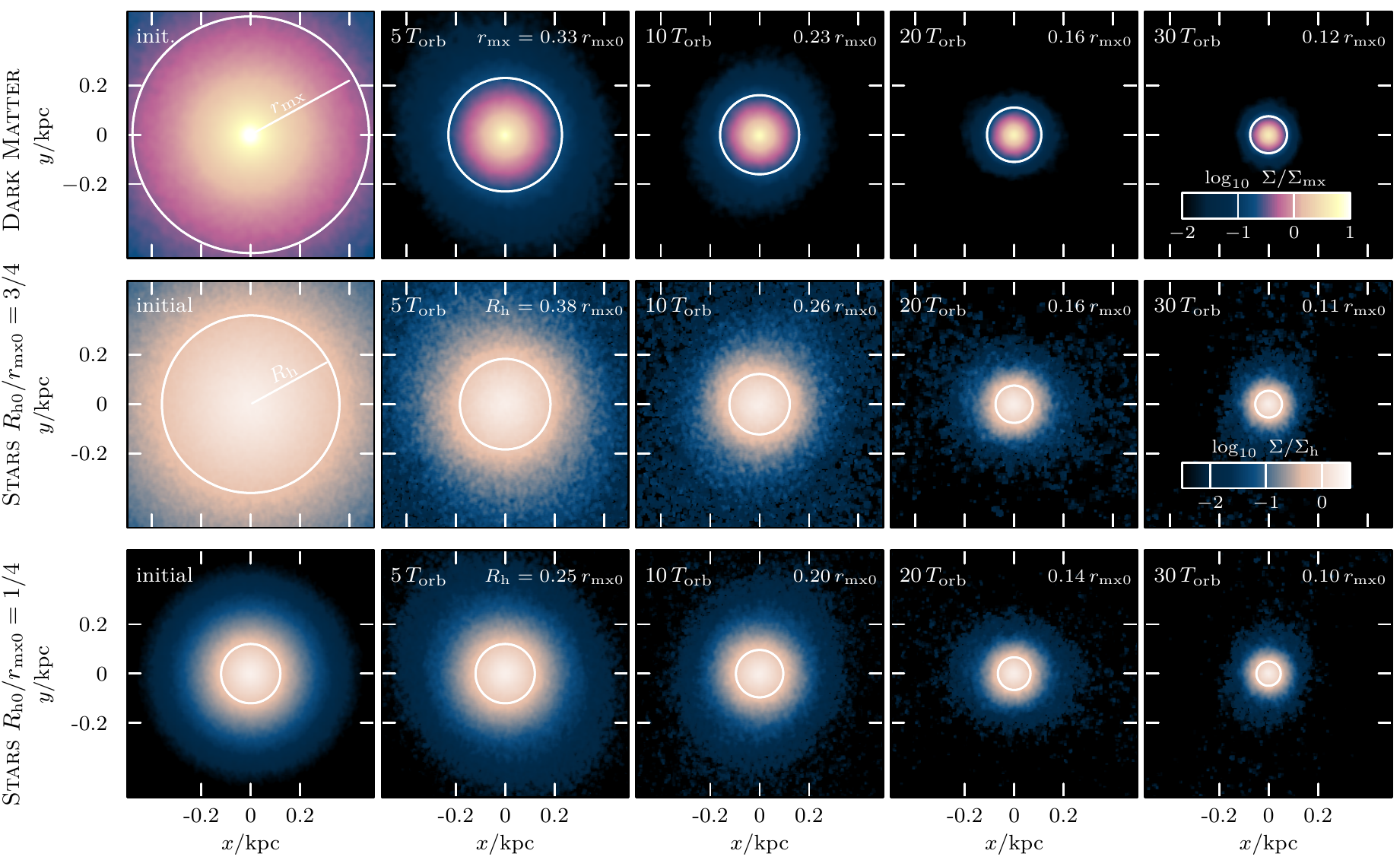}
  \caption{Tidal evolution of an NFW  subhalo (top row, model \emph{h-i}, with $M\maxzero = 10^6\,\mathrm{M_\odot}$, $r\maxzero = 0.48\,\kpc$ and $V\maxzero= 3\,\kms$)
    and two embedded exponential stellar components of different initial size (middle and bottom row for $\Rhzero/r\maxzero = 3/4$, model \emph{s-i}, and $\Rhzero/r\maxzero =1/4$, model \emph{s-ii}, respectively) on a $1$:$5$ eccentric orbit in an isothermal potential. Each panel corresponds to a simulation snapshot taken at apocentre and shows the projected surface density. The surface density is normalised by the instantaneous mean density enclosed within $\rmax$ (or $\Rh$). As tides strip the subhalo, its characteristic size $\rmax$ decreases, and the relative change in size decreases with subsequent pericentre passages: the tidal evolution slows down and a stable remnant state is asymptotically approached. Similarly, the half-light radii of embedded stellar components decrease during tidal stripping. Crucially, for the later stages of the tidal evolution, the size of the half-light radius $\Rh$ appears to follow closely the characteristic size $\rmax$ of the dark matter subhalo they are embedded in, independent of their initial extent. }
\label{fig:overview_brightness_evol}
\end{figure*}

\section{Results}
\label{sec:results}

\subsection{Tidal evolution of NFW subhalos}

\begin{figure}
 \centering
  \includegraphics[width=8.5cm]{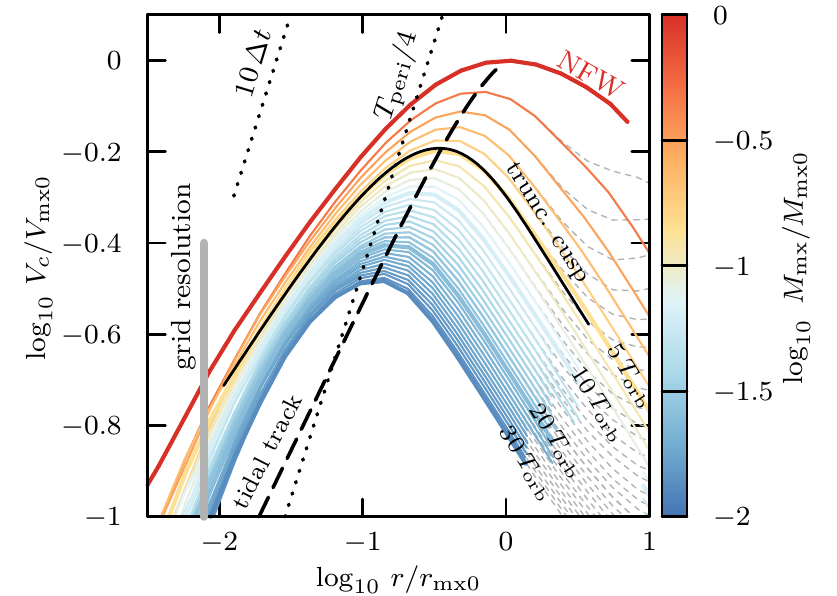}
\caption{Circular velocity profiles $\Vc(r)$ of the subhalo shown in Fig.~\ref{fig:overview_brightness_evol} (model \emph{h-i}). Each curve corresponds to a snapshot at a subsequent apocentric passage. As tides strip the subhalo, the characteristic size, $\rmax$, and velocity, $\vmax$, of the subhalo decrease along well-defined tidal tracks (dashed black curve). The distance between subsequent circular velocity curves appears to decrease with decreasing remnant bound mass $\Mmax/M\maxzero$ (see colour-coding), and tidal evolution virtually ceases once a remnant state is reached where the characteristic crossing time of the subhalo, $\tmax = 2 \pi \rmax / \vmax$, is determined by the orbital time at pericentre, $\tmax\approx \tperi/4$. For reference, the grid size of the particle mesh code is shown using a grey vertical line, and a time scale corresponding to 10 times the simulation time step $\Delta t$ is shown with a dotted black line.}
\label{fig:Vcirc}
\end{figure}

To illustrate some of the general features of the evolution of the stellar and dark matter components of NFW subhalos, we discuss here the results of a simulation of a system on an orbit with a pericentre-to-apocentre ratio of $1$:$5$. The example subhalo chosen has an initial mass of $M\maxzero=10^6\,\Msol$, an initial scale radius of $r\maxzero=0.47\,\kpc$, and an initial maximum circular velocity of $V\maxzero = 3.0\,\kms$. We will refer to this model as halo \emph{h-i}. The apocentre and pericentre radii are $200\,\kpc$ and $40\,\kpc$, respectively, and $T\maxzero/\tperi \sim 0.9$. 

We assume that the initial stellar component may be modelled as exponential spheres, 
\begin{equation}
 \rho_\star(r) = \rho_{\star 0} ~ \exp\left( -r/r_{\star s} \right)~,
\end{equation}
where $\rho_{\star 0}$ is the initial central stellar density and $r_{\star \mathrm{s}} \approx \Rh/2$ is the radial scalelength, with $\Rh$ denoting the projected stellar half-light radius.

We use the same simulation to follow the evolution of two independent stellar tracers with different degrees of radial segregation relative to the dark matter. In one tracer, the initial star-to-dark matter radial segregation is $R_\mathrm{h0} / r\maxzero = 3/4$; the second tracer is more deeply embedded inside the halo, with  $R_\mathrm{h0} / r\maxzero = 1/4$. The projected density profiles of these two systems, normalised to the same central density value, are shown in Fig.~\ref{fig:initial_final_profs} with solid red and green lines, respectively. The dark matter is shown as well, with a blue line. (Note that the relative normalization of the density is arbitrary, as the stellar components are assumed to be ``massless''.) 
We will refer to these two example stellar tracers as \emph{s-i} and \emph{s-ii}, for the extended and the deeply embedded tracer, respectively.

Figure \ref{fig:overview_brightness_evol} shows snapshots of the dark matter (top row) and of each of the two embedded stellar tracers (middle and bottom rows) in the initial conditions (left-most column), and at different apocentric passages, chosen after $5$, $10$, $20$ and $30$ orbits. 

As discussed in Sec.~\ref{SecTidalEvol}, tides gradually strip the subhalo, causing $\vmax$ and $\rmax$ to decrease, quite rapidly at first, but slowing down as the subhalo approaches the asymptotic remnant state, where $\tmax \approx \tperi/4$. The effects of tidal stripping on this orbit are quite dramatic: the mass within $\rmax$ declines to $13\%$, $6.3\%$, $2.6\%$, and $1.4\%$ of the initial value at each of the selected snapshots.

The evolution of the stellar tracers is qualitatively similar; their half-light radii decrease gradually as a result of tidal stripping, and the reduction in size decelerates as the tidal remnant stage is approached. Interestingly, the half-light radii of both stellar tracers appear to converge to the same value, despite the fact that they initially differed by a factor of $3$. After 
30 orbital periods, the difference in half-light radius is just $\sim 10\%$. Furthermore, these half-light radii (shown by solid circles in the middle and bottom rows of Fig.~\ref{fig:overview_brightness_evol}) become comparable to the characteristic radius, $\rmax$, of the dark matter (solid circles in the top row of Fig. \ref{fig:overview_brightness_evol}).

This is apparent in Fig.~\ref{fig:initial_final_profs}, where the dashed lines show the density profiles after 10 orbital periods. Solid circles on each curve mark the current value of the half-light radius. At the final time the radii are quite similar (as highlighted by the grey band), as discussed above. The total mass lost is, however, quite different. Only $\sim 6\%$ of the initial characteristic dark matter mass remains bound at that time, compared with $L/L_0 \approx 0.24$ and $0.01$ for the more and less deeply embedded stellar component, respectively.

This discussion foretells some of our main results: ``tidally limited'' systems, regardless of initial size, converge to a size set by the characteristic radius of the remnant subhalo. As we shall see below, these remnants also converge to the same velocity dispersion, which itself traces the characteristic velocity of the remnant. This implies that, in the ``tidally limited'' regime, the properties of the stellar component are poor indicators of the initial mass/size/velocity of its progenitor: stars, unlike the dark matter, follow no unique ``tidal tracks'' as they are stripped.

\begin{figure}
  \includegraphics[width=8.5cm]{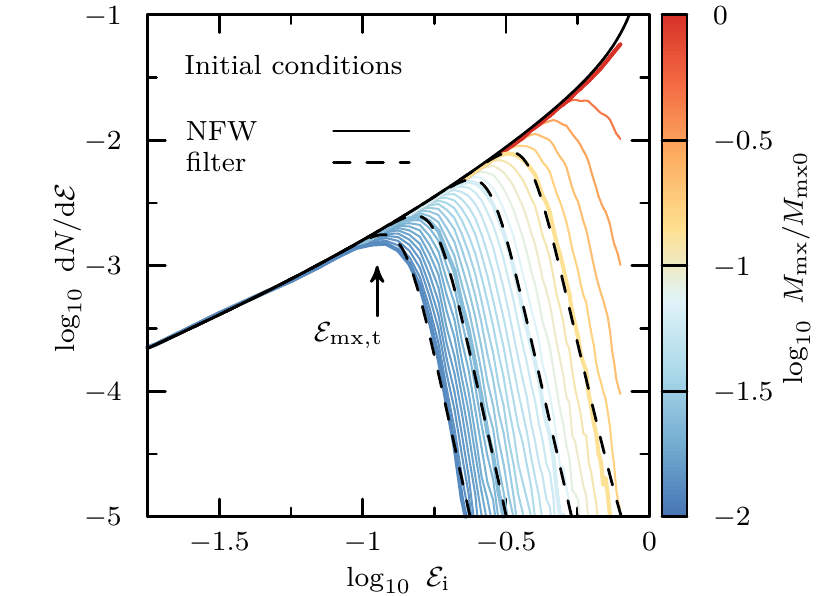} \caption{Initial binding energy distribution of particles that remain bound to the subhalo after subsequent pericentric passages (model \emph{h-i}). Initial energies are measured relative to the potential minimum, $\E = 1 - E/\Phi_0$, where for NFW halos $\Phi_0 \approx -4.63 \, \vmax^2$ (for the $N$-body realisation of an NFW halo used in this work, exponentially truncated at $10\,r_\mathrm{s}$, see Sec.~\ref{sec:subhalo-model}, we find $\Phi_0 \approx- 4.32\,\vmax^2$).  The characteristic bound mass fraction, $\Mmax/M\maxzero$, of the remnant is used for colour-coding. As tides strip the system, the remnant consists of particles of increasingly bound initial energies. The maximum of the initial energy distribution corresponding to a specific bound remnant is denoted by $\E_\mathrm{mx,t}$ and indicated for the most-stripped remnant using an arrow. The empirical filter function (Eq.~\ref{eq:DM_filter}), applied to the initial NFW energy distribution, is shown for the snapshots of Fig.\ref{fig:overview_brightness_evol}, and matches well the initial energy distribution of the bound particles. In Fig.~\ref{fig:appendix:ecc}, we show equivalent results for subhalos on circular orbits, as well as radial orbits with pericentre-to-apocentre ratios of $1\rt10$ and $1\rt20$. }
\label{fig:E_DM_ICs}
\end{figure}

\subsection{Dark matter evolution}
\label{sec:results_DM}

As discussed in the previous section, the evolution of stars and dark matter are closely coupled. We study the evolution of the dark matter in a bit more detail next, before discussing the evolution of the stellar components in Sec.~\ref{sec:results_stars}.

\subsubsection{Circular velocity profile}

As discussed by \citet{EN21}, tides lead to a gradual change in the shape of the mass profile of an NFW subhalo as well as to a reduction in its characteristic size and circular velocity. We show this in Fig.~\ref{fig:Vcirc}, where we plot the evolution of the circular velocity profile $\Vc(r) = \left[GM(<r)/r\right]^{1/2}$ of the subhalo shown in Fig.~\ref{fig:overview_brightness_evol}.

Curves are colour coded according to the remaining  bound mass fraction, $\Mmax/M\maxzero$. The subhalo evolves from an initial NFW profile towards a density  profile well-described by an exponentially truncated cusp (Eq.~\ref{eq:TruncCusp}) 
with corresponding circular velocity profile,

\begin{equation}
V_\mathrm{asy}(r) = V_\mathrm{cut} \left[\frac{1 - (1+r/r_\mathrm{cut}) \, e^{-(r/r_\mathrm{cut})} } {r/r_\mathrm{cut}}  \right]^{1/2},
\end{equation}
with $r_\mathrm{cut} \approx 0.56\,\rmax$, $V_\mathrm{cut} = \sqrt{G M_\mathrm{asy}/r_\mathrm{cut}} \approx 1.83\, \vmax$ and total mass $M_\mathrm{asy} = 4 \pi r_\mathrm{cut}^3 \rho_\mathrm{cut}$.  

The distance between adjacent circular velocity curves decreases with decreasing remnant mass, as the subhalo asymptotically approaches the asymptotic remnant state, where its crossing time is roughly a quarter of the host halo crossing time at pericentre, $\tmax \approx \tperi/4$. 

\begin{figure}  
  \includegraphics[width=8.5cm]{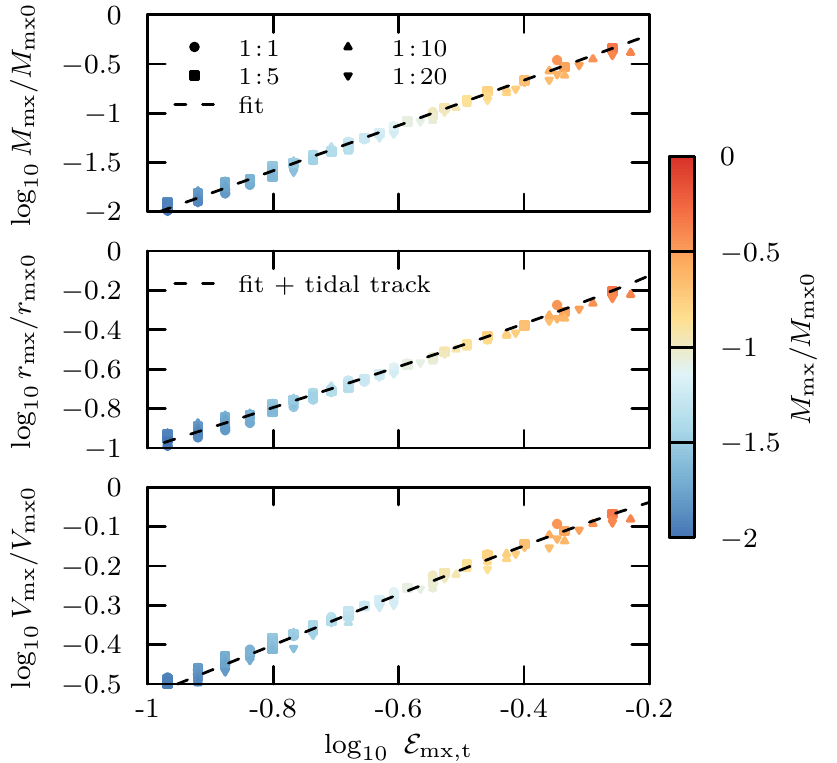}
\caption{The top panel shows the tidal truncation energy scale $\E_\mathrm{mx,t}$ in the initial conditions (see arrow in Fig. \ref{fig:E_DM_ICs}) as a function of bound mass fraction $\Mmax/M\maxzero$ of the relaxed system, which may be well approximated by a power law (for $1/100 < \Mmax/M\maxzero < 1/3$, fit Eq.~\ref{eq:Mmx_to_E}, dashed curve). The middle and bottom panels show the corresponding correlations for the characteristic size, $\rmax/r\maxzero$, and characteristic velocity, $\vmax/V\maxzero$, of the remnant, respectively. The dashed curves in the middle and bottom panel are obtained by combining the fit (Eq.~\ref{eq:Mmx_to_E}) with the tidal tracks (Eq.~\ref{eq:EN21_tracks}). Results obtained from subhalo models on orbits with different pericentre-to-apocentre ratio $r_\mathrm{peri}\rt r_\mathrm{apo}$ are shown using different symbols (for details, see Appendix~\ref{appendix:eccentricity}).}
\label{fig:E_DM_ICs-Mmx}
\end{figure}

\begin{figure}
  \includegraphics[width=8.5cm]{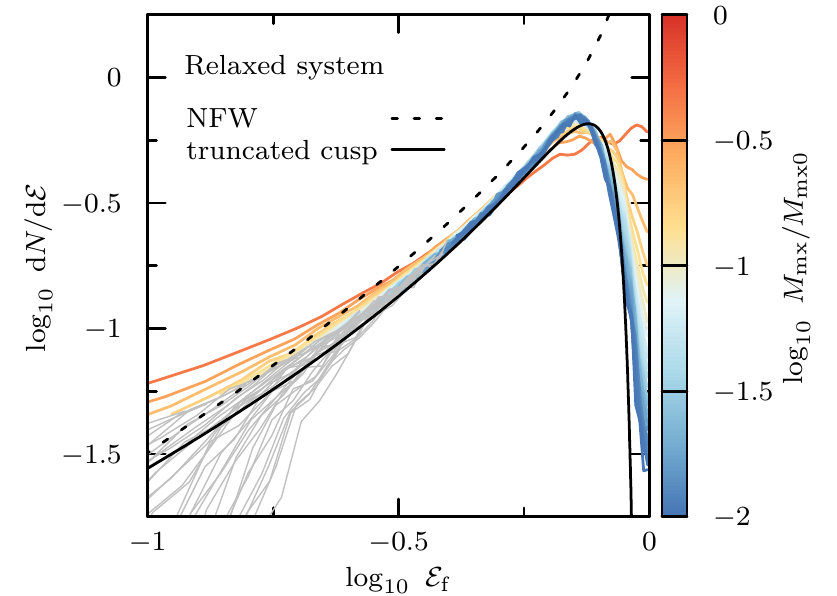}
  \caption{Energy distribution of the subhalo remnant at subsequent apocentric passages (model \emph{h-i}). Energies are defined as $\E=1 - E/\Phi_0$, with $\Phi_0 \approx - 3.35\, \vmax^2$ as expected for exponentially truncated NFW cusps.    Curves are colour coded by the bound mass fraction, and normalized to the same total mass, for ease of comparison. For remnant masses of $\Mmax/M\maxzero \lesssim 1/10$, the shape of the energy distribution evolves only weakly, as the density profile converges to its asymptotic shape. Note that the energy distribution as measured in the presence of the host halo tidal field and extra-tidal material does deviate from that of an (isolated, isotropic) exponentially truncated cusp, shown as solid black line. Energies likely compromised by numerical resolution (i.e. $E < \Phi(8 \Delta x)$) are shown in grey.}
\label{fig:energy_convergence}
\end{figure}

\begin{figure}
 \centering
 \includegraphics[width=8.5cm]{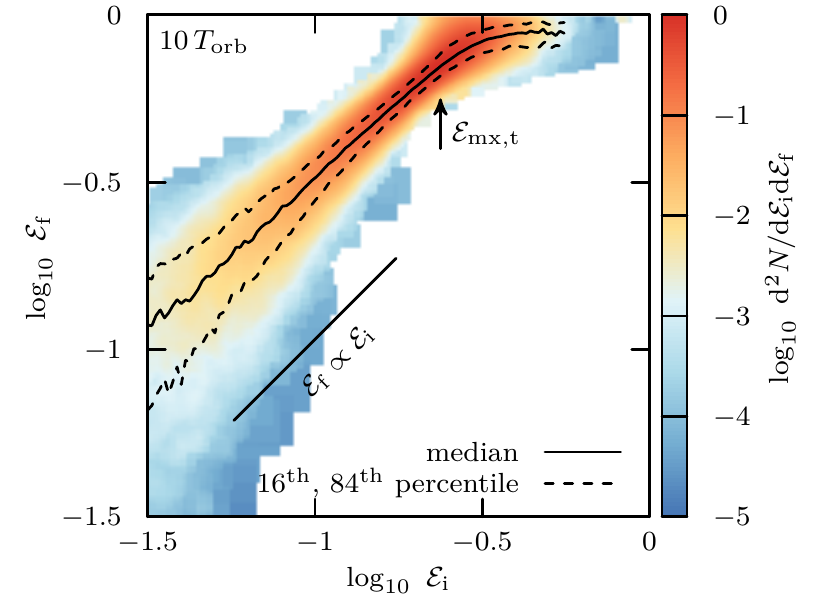}
 \caption{Initial ($\E_\mathrm{i}$) vs ``final'' ($\E_\mathrm{f}$) energy after 10 orbital periods for the bound remnant shown in Fig.~\ref{fig:overview_brightness_evol} (model \emph{h-i}). The colour-coding corresponds to the number of bound particles in each $\{\E_\mathrm{i},\E_\mathrm{f}\}$ pixel. For particles with binding energies below the tidal energy-cut $\E_\mathrm{mx,t}$ (indicated by an arrow), the median relation of the mapping is approximately linear ($\E_\mathrm{f} \propto \E_\mathrm{i}$). The 16th and 84th percentiles of the scatter around the median relation are shown using dashed curves.}
\label{fig:energy_map}
\end{figure}

\subsubsection{Binding energy}
\label{SecBindEn}
The tidal evolution is particularly simple when expressed in terms of the initial binding energy\footnote{For simplicity, we compute binding energies in the rest frame of the subhalo and do not include the gravitational effects of the main halo. Recall that our analysis focusses on the subhalo remnant structure at apocentre, where the subhalo spends most of its orbital time and where the effect of the main halo is minimal.} of the subhalo particles \citep[see; e.g.,][]{Choi2009}. This is shown in Fig.~\ref{fig:E_DM_ICs}, where we show the energy distribution of particles in the remnant at various times, again coloured by remaining bound mass fraction. We define initial binding energies in dimensionless form, $\E$, using the  potential minimum $\Phi_0$ of the initial subhalo (this is a well-defined quantity for  an NFW profile, which has a finite central escape velocity), 
$\Phi_0 \equiv \Phi(r=0)= - 4\pi G \rho_\mathrm{s} r_\mathrm{s}^2 \approx - 4.63\, \vmax^2$,

\begin{equation}
  \label{EqEnDef}
 \E \equiv 1 - E/\Phi_0,
\end{equation}
and $E = v^2/2 + \Phi(r)$.  The most-bound state is $\E = 0$, and the boundary between bound and unbound lies at $\E = 1$.  NFW energy distributions are well approximated by a power-law for $\E \rightarrow 0$. The definition used in Eq.~\ref{EqEnDef} is particularly appropriate when describing highly-stripped systems, where the least bound particles are the most likely to be stripped away by tides.

Indeed, as shown in  Fig.~\ref{fig:E_DM_ICs}, as tides strip the subhalo the energy distribution is gradually trimmed off\footnote{ The angular momentum of particles plays only a subordinate role in determining whether a particle gets stripped, as discussed in Appendix~\ref{appendix:angmom}.}, leaving only the initially most bound particles in the final remnant \citep[see also ][]{Drakos2020,stucker21,Amorisco2021}. The truncation is quite sharp, and may be approximated by a ``filter function'',
\begin{equation}
  \left. \diff N/ \diff \E \right|_\mathrm{i,t} =    {\left. \diff N / \diff \E \right|_\mathrm{i}  \over  1 + \left(a\, \E / \E_\mathrm{mx,t} \right)^{b}},
   \label{eq:DM_filter}
\end{equation}
with $a\approx0.85$, $b\approx 12$, where $ \left. {\diff N}/{\diff \E} \right|_\mathrm{i} $ denotes the initial NFW energy distribution and $\E_\mathrm{mx,t}$ the peak of the truncated energy distribution (see arrow in Fig.~\ref{fig:E_DM_ICs} marking the location of the peak for the final snapshot).

As tides strip the system, the peak energy $\E_\mathrm{mx,t}$ shifts towards more and more bound energies in the initial conditions. We therefore expect a strong correlation between the remnant bound mass, and $\E_\mathrm{mx,t}$.  This is shown in the top panel of Fig.~\ref{fig:E_DM_ICs-Mmx}, where $\E_\mathrm{mx,t}$ is seen to decrease with decreasing bound mass fraction, $\Mmax/M\maxzero$. The relation may be approximated by a power law,
\begin{equation}
\label{eq:Mmx_to_E}
 \E_\mathrm{mx,t} \approx 0.77 \, \left( \Mmax / M\maxzero \right)^{0.43},
\end{equation}
applicable for $\Mmax/M\maxzero < 1/3$.  Similarly, the middle and bottom panels of Fig.~\ref{fig:E_DM_ICs-Mmx} show the corresponding dependence of $\E_\mathrm{mx,t}$ on the characteristic size, $\rmax/r\maxzero$, and velocity, $\vmax/V\maxzero$, of the remnant. We show in Appendix~\ref{appendix:eccentricity} that the empirical description of Eq.~\ref{eq:DM_filter} and Eq.~\ref{eq:Mmx_to_E} also applies to subhalos on circular orbits, as well as to radial orbits with pericentre-to-apocentre ratios of $1\rt10$ and $1\rt20$.

Similar results are obtained when using initial orbital times instead of binding energy, as discussed in Appendix~\ref{appendix:periods}.

We have so far described the remnant using the distribution of {\it initial} binding energies. Of course, as mass is lost, the remnant relaxes and binding energies change. In particular, as the shape of the remnant density profile converges to an exponentially truncated cusp the energy distribution of the remnant converges to a well-defined shape depending only on the current potential minimum, $\Phi_0(t)$.  For an exponentially truncated cusp (Eq.~\ref{eq:TruncCusp}), the potential may be written as
\begin{equation}
 \Phi_\mathrm{asy}(r) = \Phi_\mathrm{0,asy} ~ {1-e^{-r/r_\mathrm{cut}}  \over r/r_\mathrm{cut} },
 \label{eq:Phi_Exp_Cusp}
\end{equation}
with $\Phi_\mathrm{0,asy} = - G M_\mathrm{asy}/ r_\mathrm{cut} \approx  - 3.35\, \vmax^2$. 

The energy distribution of an isolated, exponentially truncated NFW cusp with isotropic velocity dispersion may be computed analytically, and is shown by a black curve in Fig.~\ref{fig:energy_convergence}. As is clear from this figure, the remnant final energy distribution quickly approaches that of an exponentially truncated cusp. In contrast to the initial NFW profile (shown, for reference, as a dashed curve with arbitrary normalisation), which has most mass at low binding energies, the exponentially truncated cusp has a well defined peak energy, beyond which the energy distribution drops steeply.

Once the subhalo has been stripped to less than 10 per cent of its initial mass, the energy distribution corresponding to an exponentially truncated density cusp provides a good description for the energy distribution measured for the N-body remnant. Deviations at small values of $\E_\mathrm{f}$ are expected due to limitations in  numerical resolution introduced by the finite code grid size, $\Delta x$; energy values corresponding  to $E < \Phi(8 \Delta x)$ are shown in grey in Fig.~\ref{fig:energy_convergence}.

\begin{figure}
\centering
  \includegraphics[width=8.5cm]{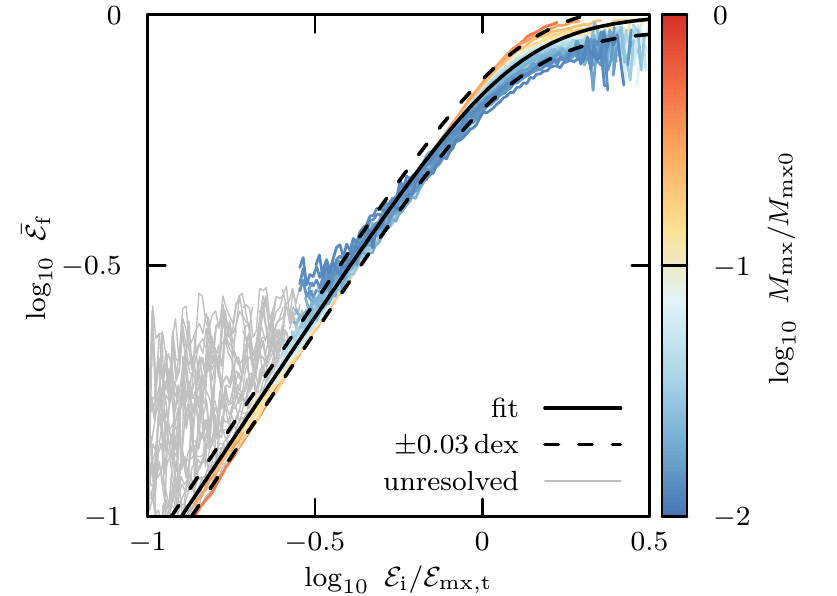}   
  \caption{Median relation shown in Fig.~\ref{fig:energy_map} for all snapshots with bound mass fraction $\Mmax/M\maxzero > 1/100$ (see colour coding), but for initial energies, $\E_i$, scaled to the tidal truncation energy scale, $\E_\mathrm{mx,t}$ (model \emph{h-i}). The relation between final and initial energies is approximately independent of tidal mass loss in these scaled units. The fit proposed in Eq.\ref{eq:energy_map} is shown by the solid black line. 
  See Fig.~\ref{fig:appendix:ecc} for equivalent plots for subhalos on circular orbits, and on radial orbits with peri-to-apocentre ratios of $1\rt10$ and $1\rt20$.  }
\label{fig:energy_map_all}
\end{figure}

\subsubsection{Initial-to-final energy mapping}
\label{sec:energy_map}

The initial and final binding energies of particles that remain bound are closely related. We illustrate this dependence in Fig.~\ref{fig:energy_map}, which shows final energies as a function of their initial values for the subhalo highlighted in Fig.~\ref{fig:overview_brightness_evol}, after $10\, T_\mathrm{orb}$. The relation is almost linear for most energy values, except for particles initially less bound than the ``peak'' energy $\E_\mathrm{mx,t}$.

Expressing initial binding energies in terms of this peak results in a relation between initial and final energies that is almost independent of the degree of tidal stripping. This is shown in  Fig.~\ref{fig:energy_map_all}, where each curve shows the median relation between between $\E_\mathrm{f}$ and $\E_\mathrm{i}/\E_\mathrm{mx,t}$ for all of our remnants. The relation is nearly independent of the remaining bound mass, especially  for heavily stripped systems with  $\Mmax/M\maxzero<0.1$ (and independent of orbital eccentricity, as shown in Fig.~\ref{fig:appendix:ecc}).
The median behaviour may be approximated by a simple empirical fit,
\begin{equation}
\label{eq:energy_map}
 \bar \E_\mathrm{f} =  \left[ 1+ (c\,\E_\mathrm{i} / \E_\mathrm{mx,t} )^{d} \right]^{1/d},
\end{equation}
with $c=0.8$ and $d=-3$.

\begin{figure}
 \centering
  \includegraphics[width=8.5cm]{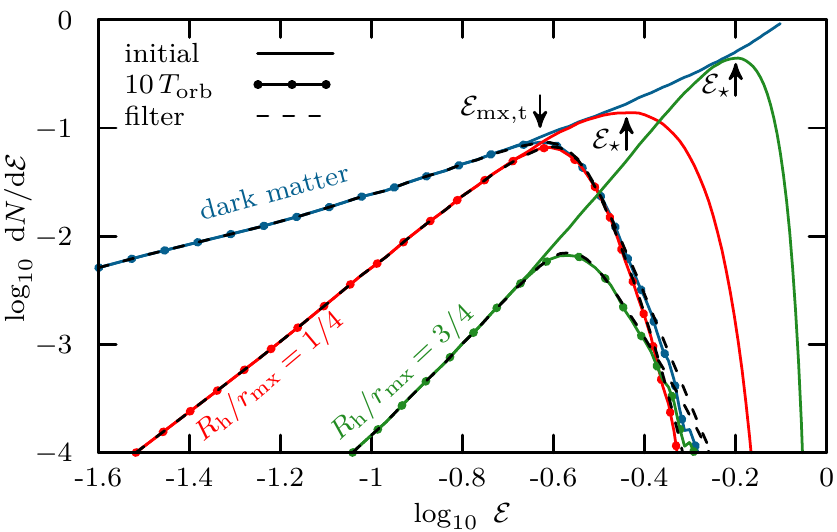}
\caption{Initial binding energy distribution of dark matter (blue, model \emph{h-i}) and stars (green and red for $\Rhzero/r\maxzero=3/4$ and $1/4$, models \emph{s-i} and \emph{s-ii}, respectively) for the simulation shown in Fig.~\ref{fig:overview_brightness_evol}. The initial distribution of all particles is shown using solid lines, 
and the distributions of those particles which remain bound after ten orbital periods are shown using dotted lines. After ten orbital periods, the tidal energy cut reaches well within both stellar distributions, and truncates them at approximately the same energy. The result of applying the empirical filter function (Eq.~\ref{eq:DM_filter}; see also Fig.~\ref{fig:E_DM_ICs}) is shown using  black dashed curves, which match the initial energy distributions of dark matter and stars that remain bound after ten orbital periods remarkably well.}
\label{fig:example_dNdE_evolved}
\end{figure}

\subsection{Evolution of the stellar component}
\label{sec:results_stars}

The results of the previous subsection enable a thorough description of the tidal evolution of the dark matter component of an NFW subhalo. Under the assumption that stars may be treated as tracers of the gravitational potential, their evolution would just follow that of a subset of the dark matter, defined mainly by their initial binding energy distribution.

Let us consider again the example of Fig.~\ref{fig:overview_brightness_evol}.  The initial energy distributions of stars and dark matter are shown with solid curves in Fig.~\ref{fig:example_dNdE_evolved}. For exponential profiles, the most bound regions  are well approximated by power-laws that peak at $\E_\star$ (a characteristic value which depends on the radial segregation of stars and dark matter) and are sharply truncated beyond $\E_\star$.

The initial energy distribution of particles that remain {\it bound} after 10 orbital periods are shown by the lines connecting circles of the corresponding colour. The black dashed curves overlapping each of these lines indicate the result of applying the ``filter function'' introduced in Sec.~\ref{SecBindEn} (Eq.~\ref{eq:DM_filter}). Note that the {\it same} filter, applied to either stars or dark matter, yields an excellent description of the initial energy distribution of the particles that remain bound after $10\, \Torb$.

This has a number of important implications. It suggests that (i) in practice the {\it only} parameter that matters when determining the probability that a given particle will remain bound (or be stripped) is its initial binding energy.  It also implies that (ii) stellar components  whose initial energy distributions peak beyond the truncation energy, $\E_\mathrm{mx,t}$, will share the same outermost energy distribution as the dark matter remnant. Finally, (iii) it suggests that the final structure of the bound stellar remnant will depend on how stars populate energies more bound than $\E_\mathrm{mx,t}$ in the initial system; in other words, the ``tidal tracks'' and final structure of the stellar remnant should depend sensitively on the stars' initial density profile and radial segregation relative to the dark matter. We explore these implications in more detail next.

\begin{figure}
 \centering
   \includegraphics[width=8.5cm]{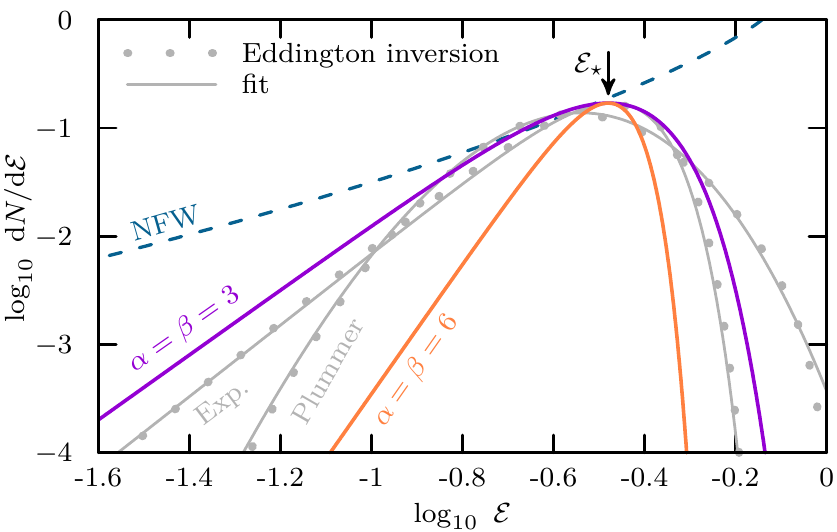}  
   \includegraphics[width=8.5cm]{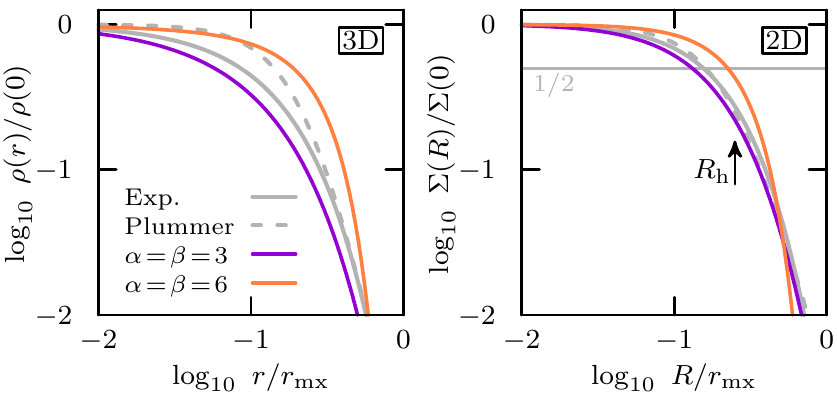}  
\caption{Density (bottom left) and surface density profiles (bottom right) corresponding to the energy distribution of Eq.~\ref{eq:this_dnde} (with $\alpha=\beta=3$ shown in magenta, and $\alpha=\beta=6$ in orange), for stellar tracers with radius $\Rhzero/r\maxzero = 1/4$ embedded in an NFW dark matter halo. For reference, the corresponding profiles for exponential and Plummer spheres are shown as grey solid and dashed curves, respectively. The top panel shows the energy distributions corresponding to the density profiles shown below. The energy distribution for exponential and Plummer spheres is computed using Eddington inversion and shown as grey dots; grey solid curves show fits of Eq.~\ref{eq:this_dnde} to these distributions, highlighting that the functional form of Eq.~\ref{eq:this_dnde} is sufficiently flexible to enable the modeling of a representative range of different stellar systems.}
\label{fig:Stellar_dNdEs}
\end{figure}

\subsubsection{Initial and final stellar remnant structure}
As mentioned above, our results suggest that the structure of the stellar remnants will be linked to its initial energy distribution and to the degree of stripping undergone by the subhalo. We adopt a flexible parameterization of the stellar energy distribution in order to examine the relation between the initial and final density profiles of the stars. In particular, we explore initial binding energy distributions of the form:
\begin{equation}
 \diff N_\star / \diff \E =  \begin{cases}
                        \E^\alpha ~ \exp \left[- (\E/\E_\mathrm{s})^\beta  \right] & \text{if}~ 0 \leq \E < 1 \\
                        0 & \text{otherwise,}
                      \end{cases}
\label{eq:this_dnde}
\end{equation}
where, as usual, $\E = 1 - E/\Phi_0$. The distribution behaves like a power-law towards the most-bound energies ($\E \rightarrow 0$), is exponentially truncated beyond some scale energy $\E_\mathrm{s}$, and it peaks at $\E_\star = \E_\mathrm{s} ~ \left( \alpha/\beta \right)^{1/\beta} < 1$.

\begin{figure}
 \centering
    \includegraphics[width=8.5cm]{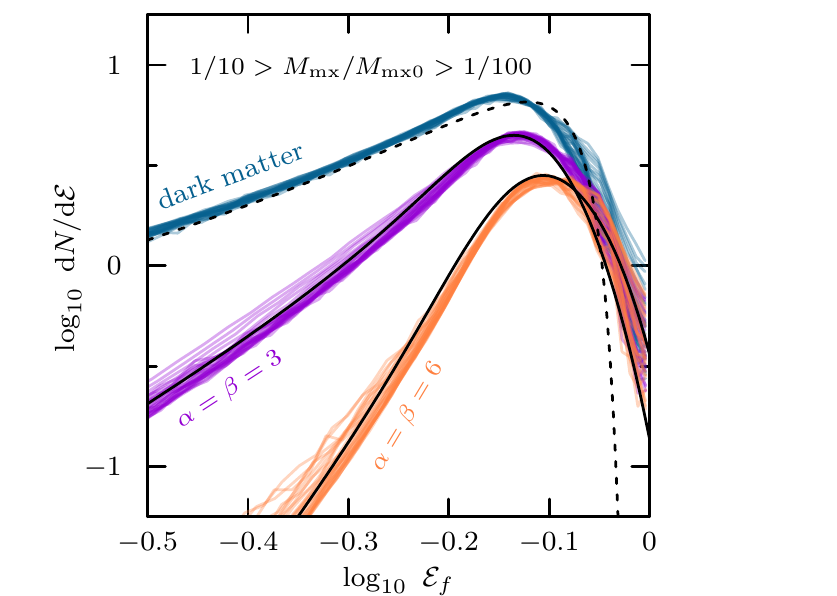}  
\caption{Energy distributions of dark matter (blue, evolved model \emph{h-i}) and stars (magenta and orange for initial $\alpha=\beta=3$ and $\alpha=\beta=6$ profiles, respectively) in the relaxed remnant system (arbitrary normalisation). Only snapshots where the tidal energy truncation reaches well into the initial energy distribution of the stars are shown.  All three energy distributions peak at approximately the same energy ($\log_{10}\,\E_\mathrm{f} \approx -0.15$). The energy distribution for an isolated, exponentially truncated cusp with isotropic velocity dispersion is shown as a black dashed curve. For the stellar models, energy distributions computed using the empirical model of Appendix~\ref{appendix:step-by-step} are shown as black solid curves.}
\label{fig:evolved_stellar_dNdE}
\end{figure}

Fig.~\ref{fig:Stellar_dNdEs} shows two examples, one with $\alpha=\beta=3$ (shown in magenta) and another with $\alpha=\beta=6$ (shown in orange).  The value of $\E_\mathrm{s}$ is chosen so that both models have the same initial 2D half-light radius, $R_\mathrm{h0} / r\maxzero= 1/4$ (isotropic solutions for different initial half-light radii are discussed in Appendix~\ref{appendix:isotropic_models}).
As $\alpha =\beta$, the scale energy and peak energy coincide; i.e., $\E_\star = \E_\mathrm{s} \approx 1/3 $. The $\alpha=\beta=3$ model spans a much wider range of energies than the $\alpha=\beta=6$ model (see top panel), and thereby also samples more energies that probe deeper into the dark matter density cusp. As a consequence, the resulting surface brightness profile of the $\alpha=\beta=3$ model has a smaller core radius (defined as the radius where the projected stellar density drops by a factor of $2$ from its central value; i.e., $\Sigma(R_\mathrm{c}) = \Sigma(0)/2$) than the $\alpha=\beta=6$ model. The corresponding 2D and 3D density profiles are shown in the bottom panels of Fig.~\ref{fig:Stellar_dNdEs}. For reference, the ratio $\Rc/\Rh$ between core and half-light radius in the initial conditions is $0.5$  and $0.9$ for  $\alpha=\beta=3$ and  $\alpha=\beta=6$, respectively. 

Many different density profiles may be approximated by varying $\alpha$ and $\beta$ in Eq.~\ref{eq:this_dnde}.  For example, a Plummer model may be approximated quite well by $\alpha=14,\beta=0.45$, whereas exponential profiles are well fit by 
$\alpha=3.3,\beta=4.0$. Indeed, the density profile for the $\alpha=\beta=3$ case is almost indistinguishable from an exponential profile over two decades in radius, as may be seen in the bottom panels of Fig.~\ref{fig:Stellar_dNdEs}.

As expected, the final and initial energy distribution of the stars are highly dependent on each other. For example, we show in
Fig.~\ref{fig:evolved_stellar_dNdE} the final  energy distributions for the $\alpha=\beta=3$ (magenta) and $\alpha=\beta=6$ (orange) stellar tracers in the ``tidally limited'' regime; i.e., after substantial mass loss. Various snapshots are shown at apocentric passages and span a large range of bound mass fraction: $1/100 < \Mmax/M\maxzero < 1/10$. Energies $\E_\mathrm{f} = 1-E/\Phi_{0}$ are normalised by their instantaneous potential minimum (approximated as $\Phi_0 \approx \Phi_\mathrm{0,asy} \approx -3.35\, \vmax^2$).
The shape of the energy distributions converges rapidly (as shown in Fig.~\ref{fig:energy_convergence} for dark matter alone), and little evolution in the energy distribution shape is seen for remnant masses smaller than $0.1\, M\maxzero$. 

The final energy distributions of the remnants may also be computed using the ``filter'' function of Eq.~\ref{eq:DM_filter} to select  bound particles from the initial conditions, combined with the initial-to-final energy mapping given by Eq.~\ref{eq:energy_map}. The result of this calculation is shown by the solid black lines in Fig.~\ref{fig:evolved_stellar_dNdE} and are seen to approximate well the results of the simulations. A step-by-step description of how to implement this calculation is given in Appendix~\ref{appendix:step-by-step}. 

\begin{figure}
 \centering
    \includegraphics[width=8.5cm]{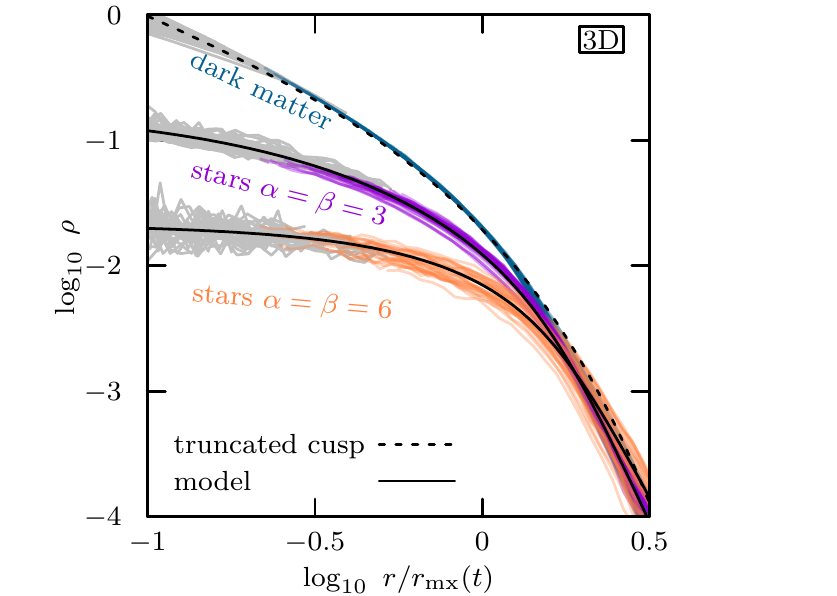} 
 \caption{Density profiles corresponding to the models discussed in Fig.~\ref{fig:evolved_stellar_dNdE} (vertical  normalisation is arbitrary). While the characteristic size of all three profiles is set by the tides ($\Rh \sim \rmax$), the inner regions of the stars retain memory of the initial profile and energy distribution. The tidally limited density profile of the NFW subhalo (evolved model \emph{h-i}) is well-described by an exponentially truncated cusp (dashed curve). For the stellar models, density profiles have been computed from the empirical energy distribution discussed in Appendix~\ref{appendix:step-by-step}, and are shown as solid black curves. Radii which are likely affected by the limited numerical resolution are shown in grey (see Sec.~\ref{sec:simcode}).  } 
 \label{fig:profiles}
\end{figure}

The different initial energy distributions give rise to different density profiles for the stellar remnants, depending on the values of $\alpha$ and $\beta$ chosen. We show this in Fig.~\ref{fig:profiles}, where we plot the 3D density profiles corresponding to the energy distributions shown in Fig.~\ref{fig:evolved_stellar_dNdE}. Radii are scaled to the instantaneous characteristic size of the underlying dark matter halo, $\rmax(t)$. All profiles are sharply truncated beyond $\rmax$, and have similar outer slopes (consistent with the findings of \citet{Penarrubia2009}, who studied tidal stripping of stellar King models deeply embedded within NFW subhalos). The inner regions, however, differ, retaining clean memory of their initial profiles. As in the initial conditions, the $\alpha=\beta=6$ remnant profile has a larger ratio of core to half-light radius ($\Rc/\Rh\sim 0.8$) than the $\alpha=\beta=3$ profile ($\Rc/\Rh\sim 0.6$).

\subsubsection{Tidal tracks}

\begin{figure}
 \centering
  \includegraphics[width=8.5cm]{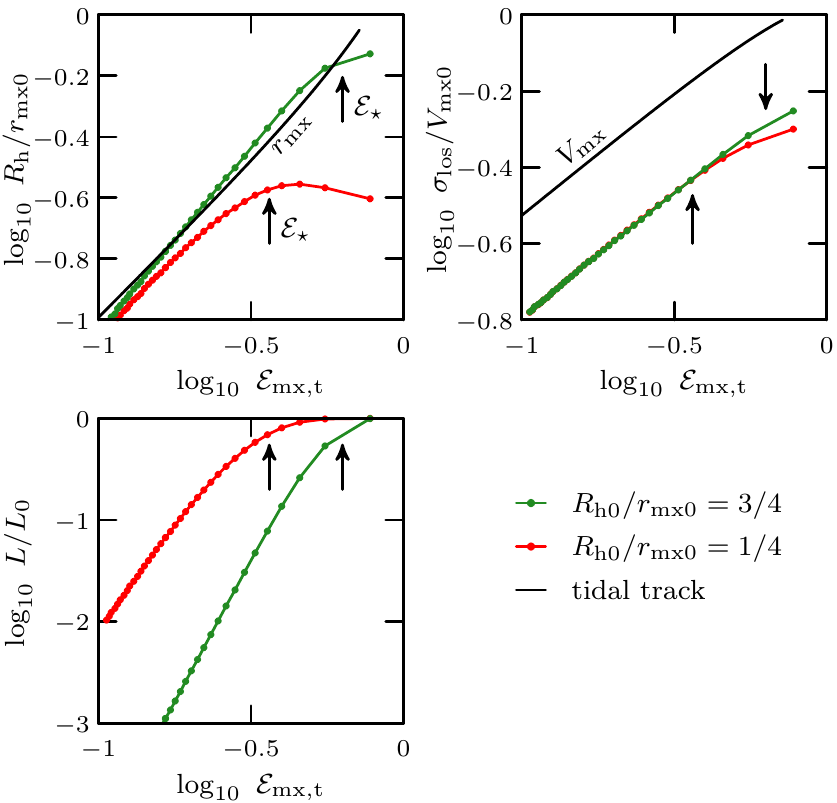}   
  \caption{Evolution of projected stellar half-light radius, $\Rh$, line-of-sight velocity dispersion, $\sigmalos$, and luminosity (stellar mass), $L$, as a function of the tidal energy cut $\E_\mathrm{mx,t}$, for the simulation shown in Fig.~\ref{fig:overview_brightness_evol} (stellar tracers \emph{s-i} and \emph{s-ii}, embedded in subhalo \emph{h-i}). Half-mass radii and velocity dispersions are normalised to the initial characteristic radius, $r\maxzero$, and characteristic velocity, $V\maxzero$, of the dark matter halo, respectively. The evolution of $\rmax$ and $\vmax$ is shown using black solid curves, and proceeds from right to left in these panels. Before the tidal energy cut reaches the characteristic energy of the stellar tracer (i.e., while $\E_\mathrm{mx,t} > \E_\star$), the stellar half-light radius and total luminosity change only slightly. Once $\E_\mathrm{mx,t} \lesssim \E_\star$, the half-light radius decreases rapidly and approaches the characteristic radius $\rmax$ of the dark matter halo.  Similarly, for $\E_\mathrm{mx,t} \lesssim \E_\star$, the line-of-sight velocity dispersion closely traces the subhalo characteristic velocity, $\sigmalos \approx \vmax / 2$, and the total luminosity of the bound stellar component drops rapidly.  }
  \label{fig:example_stellar_evol}
\end{figure}

The results presented in Fig.~\ref{fig:example_dNdE_evolved} suggest that luminosity and size of the stellar components should remain largely unaffected by tides until the truncation in energy starts to overlap the energy distribution of the stars. Using the notation illustrated in that figure, this occurs when $\E_\mathrm{mx,t}$ becomes comparable or smaller than $\E_\star$. We show this in Fig.~\ref{fig:example_stellar_evol}, where we plot the evolution of the stellar projected half-light radius $\Rh$, as well as of the central line-of-sight velocity dispersion $\sigmalos$ and luminosity (i.e., stellar mass), $L$, of the two stellar populations highlighted in the example shown in Fig.~\ref{fig:overview_brightness_evol}.

The evolution is plotted as a function of the tidal truncation energy, $\E_\mathrm{mx,t}$, which is a proxy for the total mass loss due to tides (see Fig.~\ref{fig:E_DM_ICs-Mmx}). The smaller $\E_\mathrm{mx,t}$ the higher the mass loss, so tidal evolution runs from right to left in Fig.~\ref{fig:example_dNdE_evolved}. The location of the peak $\E_\star$ for each of the two initial stellar energy distributions is indicated by the vertical arrows.

As expected, if the tidal truncation energy does not overlap the stellar energy distribution (i.e., if $\E_\mathrm{mx,t}>\E_\star$), the size of the stellar component remains largely unaffected by tides. However, once $\E_\mathrm{mx,t}$ drops below $\E_\star$, the stellar components rapidly lose mass, decrease in size, and gradually become kinematically colder. As expected, the more extended stellar component ($\Rhzero / r\maxzero = 3/4$) is more readily affected than the more deeply segregated component with $\Rhzero / r\maxzero = 1/4$.

Interestingly, once $\E_\mathrm{mx,t}$ becomes much smaller than $\E_\star$, the stellar half-light radius converges to the dark matter characteristic radius, $\rmax$, regardless of its initial value. In addition,  the velocity dispersion becomes directly proportional to the dark matter characteristic velocity, $\vmax$. In other words, the stellar components of these ``tidally limited systems'' become direct tracers of the characteristic parameters of the tidal remnant, almost regardless of their initial radial segregation \citep[see, also,][]{Kravtsov2010}.

\begin{figure}
 \centering
    \includegraphics[width=8.5cm]{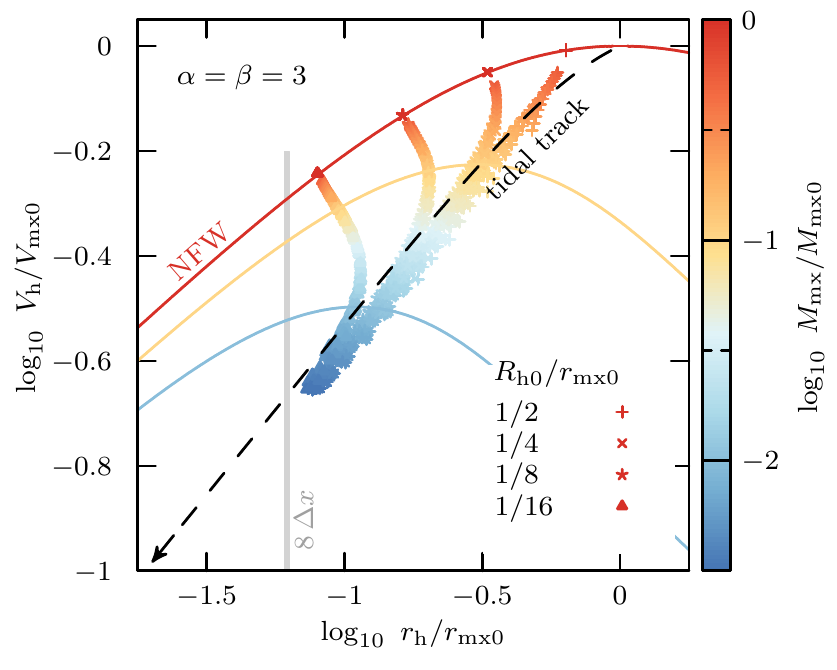} 
    \caption{Three-dimensional half-light radii, $r_\mathrm{h}$, as well as circular velocities at that radius, $V_\mathrm{h}\equiv V_\mathrm{c}(r_\mathrm{h})$ measured from the simulations described in Sec.~\ref{sec:nummethods}. The evolution of four different stellar tracers with initial segregations $\Rhzero/r\maxzero =1/2,1/4,1/8,1/16$ are shown using different symbols, which have been embedded as massless tracers in all subhalos of the \citet{EN21} set of simulations on orbits with pericentre-to-apocentre ratio $1\rt5$. 
    Symbols are colour-coded by bound mass fraction. The  circular velocity profiles  of the remnants for $\Mmax/M\maxzero=1,1/10$ and $1/100$ are shown using coloured solid lines. Extended stellar tracers rapidly decrease in size once tides truncate the underlying subhalo, but more segregated tracers first expand slightly before being truncated by the tides. The parameters of embedded stellar systems stay always close or above the dark matter subhalo tidal track (black dashed curve).}
\label{fig:stellar_track_overview}
\end{figure}

This is a general result, which applies to all systems in the ``tidally limited'' regime. We illustrate this in Fig.~\ref{fig:stellar_track_overview} where we plot the ``tidal tracks'' of  $\alpha=\beta=3$ systems with different initial radial segregation; i.e., $\Rhzero/r\maxzero =1/2$, $1/4$, $1/8$,  and $1/16$, each shown with different symbols (Appendix~\ref{appendix:tidal_tracks} shows the same tidal tracks in units of remnant bound mass fraction $\Mmax/M\maxzero$). As the subhalo gets stripped, its characteristic radius and velocity ($\rmax$ and $\vmax$) follow the track indicated by the dashed line. The 3D half-light radius and velocity dispersion of the stellar components (the latter multiplied by $\sqrt{3}$ to convert it to a circular velocity) evolve as well, generally getting smaller and colder as stripping progresses.

Only the most embedded tracers ``puff up'' initially, increasing their size\footnote{Appendix~\ref{appendix:e_vs_r} discusses in more detail the slight expansion of stellar tracers before the onset of strong tidal stripping, as well as the drop in velocity dispersion even before a reduction in size is observed.} as they get kinematically colder, but the net increase is small. Interestingly, once the stellar component parameters reach the tidal track of the dark matter remnant, they follow it closely. Qualitatively, this result applies to all the values of $\alpha$ and $\beta$ we have tried, and it therefore appears of general applicability: the size and velocity dispersion of a tidally-limited stellar system traces closely the $\rmax$ and $\vmax$ of the dark remnant\footnote{For stellar models with initial $\alpha=\beta=3$ and $\alpha=\beta=6$ energy distributions, in the tidally limited regime, we measure $\Rh \approx 0.93\,\rmax$, $\sigmalos \approx 0.70\,\vmax$, and  $\Rh \approx 1.26\,\rmax$, $\sigmalos \approx 0.76\,\vmax$, respectively, see Appendix~\ref{appendix:tidal_tracks}.}. 

This result has a couple of interesting implications. One is that there is no unique ``tidal track'' describing the evolution of a stellar system in the size-velocity plane, as had been suggested by \citet{Penarrubia2008}, i.e., the tidal track depends on the initial distribution of stars within the subhalo. With hindsight, it is easy to trace that conclusion to the fact that those authors tested a single stellar model (a King profile) and a rather limited range of radial segregation. In fact, Fig.~\ref{fig:stellar_track_overview} shows clearly that the same stellar tidal remnant may be produced from very different initial radial segregations, each following different tracks.

A second implication is that the tidal track of a subhalo (i.e., the dashed line in Fig.~\ref{fig:stellar_track_overview}) sets a firm lower limit on the velocity dispersion of an embedded stellar remnant of given size. (Or, equivalently, an upper limit to the size of an embedded system of given velocity dispersion.)

We discuss next an application of this argument to the interpretation of the sizes and velocities of dwarf galaxies in the Local Group.

\begin{figure*}
 \centering
    \includegraphics[width=8.5cm]{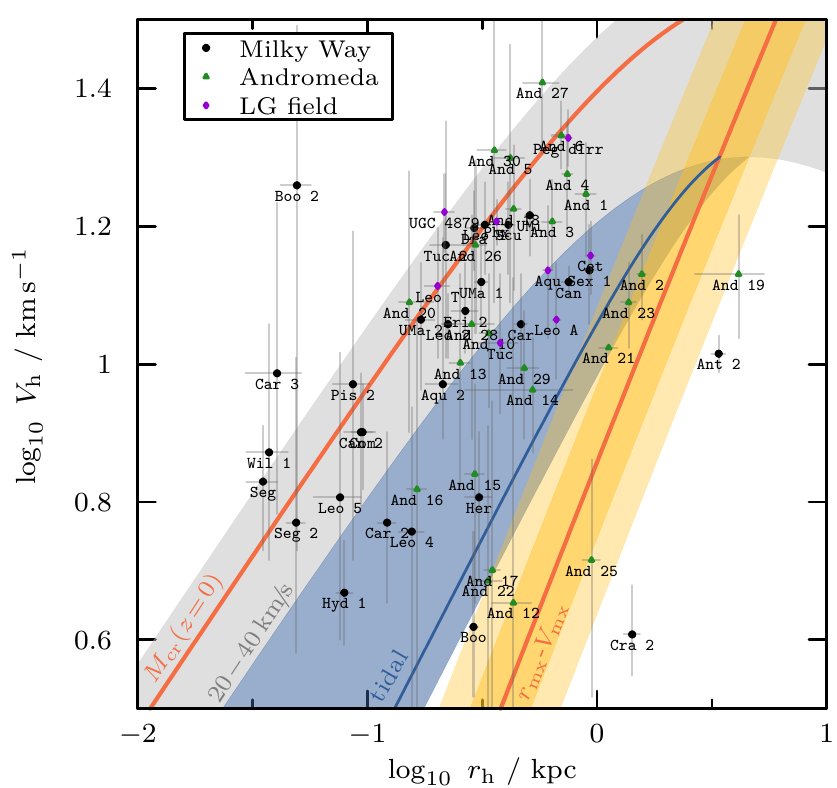}     \includegraphics[width=8.5cm]{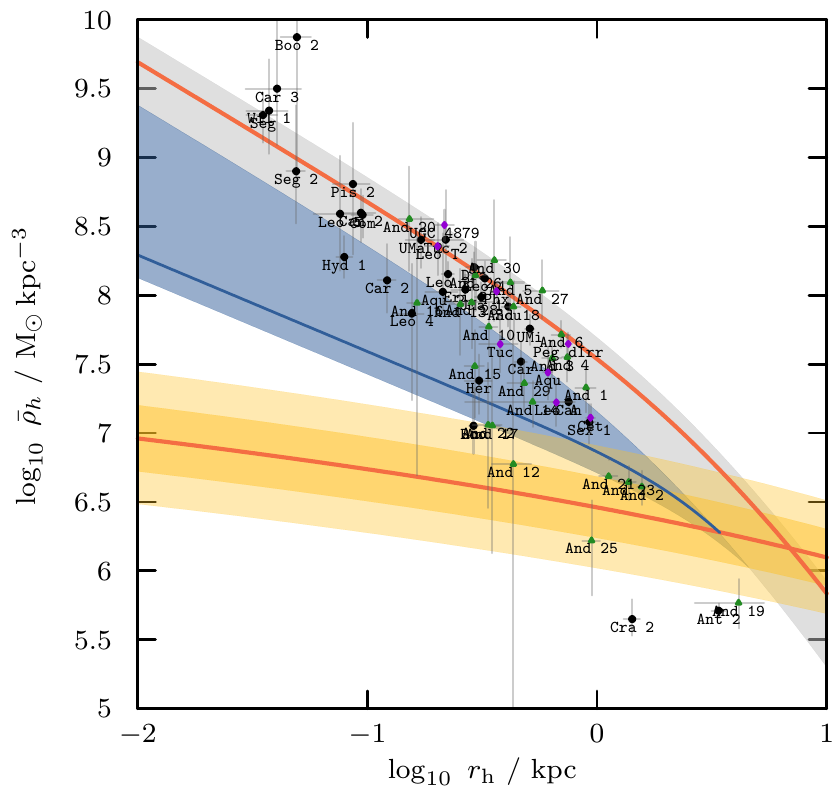} 
    \caption{Three-dimensional half-light radii, $r_\mathrm{h}$, and circular velocities $V_\mathrm{h}$, for $L < 10^7\,\mathrm{L_\odot}$ Milky Way- and Andromeda satellites, as well as Local Group field galaxies (left panel). Observational values are estimated using the \citet{Wolf2010} mass estimator ($r_\mathrm{h} \approx 4/3\,\Rh$, $V_\mathrm{h} \approx \sqrt{3} \sigma_\mathrm{los}$) from the half-light radii and velocity dispersions listed in \citet{McConnachie2012} (version January 2021, and updated velocity dispersions for Antlia 2, Crater 2 \citep{Ji2021}, Tucana \citep{Taibi2020}, And 19 \citep{Collins2020} and And 21 \citep{Collins2021}). The cosmological $\rmax$-$\vmax$ (i.e., mass-concentration) relation at redshift $z=0$ is shown using a solid red curve, with yellow bands corresponding to $0.1\,$dex scatter in concentration \citep{Ludlow2016}. A grey band indicates NFW halos with maximum circular velocity in the range $20\,\kms < \vmax < 40\,\kms$, and delineates the regions where all isolated dwarfs should lie \citep[see,e.g.,][]{Fattahi2018}. For reference, the circular velocity profile of a mean-concentration halo with a mass equal to the critical mass for hydrogen cooling $M_\mathrm{cr}(z=0) \approx 7.5 \times 10^9\,\Msol$ \citep{Benitez-Llambay2020} is shown in red. A tidal track from the lowest-mass subhalo of that band (blue solid curve) puts a lower bound on the region where tidally-stripped dwarf galaxies could be found. Most satellite galaxies shown fall in either the grey shaded or blue shaded region and are consistent with these constraints. Dwarfs outside the blue and gray regions are not easily explained in this framework. The panel on the right shows an alternative view of the same data, but cast in terms of mean density $\bar \rho_\mathrm{h}$ within the half-light radius $r_\mathrm{h}$. The dwarf galaxies that fall below the grey and blue shaded regions have densities too low to be consistent with stripped NFW subhalos. See text for further discussion.}
\label{fig:dwarfs_vs_ludlow}
\end{figure*}

\subsection{Application to Local Group dwarfs}

We now use the results of the previous subsections to interpret available structural data on Local Group dwarfs, in the context of LCDM. As discussed in Sec.~\ref{SecIntro}, the assumption of a minimum mass for galaxy formation, such as that expected from the hydrogen cooling limit, together with the mass-concentration relation expected for NFW halos in LCDM, yield strong constraints on the size and velocity dispersion of dwarf galaxies.

Following \citet{Fattahi2018}, the grey band in Fig.~\ref{fig:dwarfs_vs_ludlow} indicates the circular velocity profiles of NFW halos with $\vmax$ in the range $20$-$40$ km/s, the virial mass range expected to host all isolated dwarfs with $M_\star<10^7\, M_\odot$. The solid orange line indicates the $\rmax$-$\vmax$ relation expected from the LCDM mass-concentration relation \citep{Ludlow2016}. The accompanying error band delineates the one- and two-sigma scatter inferred from cosmological N-body simulations.

For dSphs, the velocity dispersion of the stars may be used to estimate the circular velocity, $\Vh$, at the 3D stellar half-light radius, $\rh$, using the \citet{Wolf2010} mass estimator\footnote{In contrast to the \citet{Walker2009} and the \citealt{EPW18} mass estimators, which are calibrated to estimate the mass enclosed within the 2D projected half-light radius $\Rh$ (and within $1.8\,\Rh$, respectively), the \citet{Wolf2010} estimator is tailored to estimate the mass enclosed within the 3D deprojected half-light radius, which facilitates the comparison to circular velocity curves.}.  While not exact, we expect these estimates to be accurate to about  0.1 dex \citep[see also][]{Walker2009, Campbell2016, EPW18}. Dwarfs unaffected by tides should lie on the grey area if they follow LCDM predictions. Encouragingly, Local Group {\it isolated } (field) dwarfs, shown by magenta symbols in Fig.~\ref{fig:dwarfs_vs_ludlow}, seem to be in reasonable agreement with that prediction.

Systems below the grey band are, in the LCDM context, typically interpreted as satellites whose velocity dispersions and sizes have been affected by tides, bringing them {\it below} the grey band. Indeed, no dwarf with $M_\star <10^7\, M_\odot$ should fall {\it above}  the grey band, according to LCDM. The ones that seem to deviate from that precept, such as Bootes 2 and Carina 3, have fairly large uncertainties in their circular velocity estimates, mainly due to the extremely small number of stars with measured radial velocities. Furthermore, \citet{Ji2016_Bootes2} argue that the velocity dispersion of Bootes 2 is likely over-estimated due to the presence of  binaries, and that the reported velocity dispersion is best seen as an upper limit. It remains to be seen whether these galaxies are actually in conflict with LCDM once their velocity dispersions are better constrained.

Galaxies below the grey band are all satellites, a requisite for a tidal interpretation of their location in the size-velocity plane. However, tides cannot shift them to arbitrarily low velocities without affecting their size, as shown by our discussion of the tracks in Fig.~\ref{fig:stellar_track_overview}. Indeed, the tidal track of the halo with the minimum mass needed to form a galaxy should provide firm limits on the size and velocity of equilibrium ``tidally limited'' dwarfs. This may be seen as a lower limit on the velocity at given size, or, equivalently, as an upper limit to the size given a velocity.

This limit is indicated by the blue line in Fig.~\ref{fig:dwarfs_vs_ludlow}, which outlines the tidal track of a $\vmax=20$ km/s halo of average concentration.
This restricts the loci of tidally limited dwarfs to the region shaded in blue: no dwarf should populate the region below the blue line.

Interestingly, there are a number of satellites that seem to violate this condition. These are galaxies of unusually low velocity dispersion for their size, like the MW ``feeble giant'' satellites Crater 2 and Antlia 2, as well as the M31 satellites And 19, And 21 and And 25. Our results imply that tidal effects are not a viable explanation for these galaxies in the context of LCDM. 

These conclusions are consistent with those of \citet{Sanders2018}, who reported ``tension'' when trying to account for the size of Crater 2 in tidally limited NFW halos, and argued that the tension may be alleviated if the progenitor subhalo had a sizable constant-density ``core''. \citet{Sanders2018} reached these conclusions considering also flattened models in their analysis, which suggests that angular momentum plays a subdominant role in determining the remnant structure of dispersion-supported tidally stripped systems, see Appendix~\ref{appendix:angmom}.

On the other hand, \citet{Frings2017} reports simulations that can apparently produce satellites as large and as kinematically cold as Crater 2. However, these results are likely affected by numerical limitations. For example, the closest ``Crater 2 analogue'' that they report corresponds to a halo whose maximum circular velocity has been reduced to nearly $1/10$th of its original value. As discussed by \citet{EN21}, adequately resolving the tidal remnant in that case is extremely difficult, and would require many more particles and much better spatial resolution than used by \citet{Frings2017}. In addition, some of the \citet{Frings2017} satellites have density profile cusps somewhat shallower than NFW, hindering a detailed comparison with our results.

Finally, \citet{Amorisco2019} argued that transients induced by tides may be the cause of the large radii and low velocity dispersions of these objects. This interpretation could in principle be checked by looking for other signs of ongoing tidal disturbance, such as a velocity gradient across the system \citep{Li2021}, extended tidal tails, or ``bumps'' in the surface density profile caused by escaping stars \citep{Penarrubia2009}. We have not considered such transients in the analysis presented here, which only considers the (equilibrium) structure of galaxies near apocentre.

To summarize, galaxies below the blue tidal track shown in Fig.~\ref{fig:stellar_track_overview} present a serious challenge to the tidal interpretation suggested by models based on the LCDM scenario. Possible resolutions of this challenge include: (i) revisions to the observed photometric or kinematic parameters of these galaxies, many of which are extremely challenging to study due to their extremely low surface brightness; (ii) the possibility that these galaxies are not bound, equilibrium systems, but are in the process of being tidally stripped,
(iii) that these systems formed in halos with masses substantially below the hydrogen cooling limit,
or (iv) that the dark matter halo of these galaxies deviates from the NFW shape, perhaps signalling the effects of baryons on the inner cusp of the halo even in ultrafaint galaxies\footnote{We note that some of the ultra-faints, if tidally limited, may have been more luminous, since stripping may also affect the luminous component. This may leave a discernible imprint in the location of these galaxies in the mass-metallicity plane.},
or, perhaps more intriguingly, effects associated with the intimate nature of dark matter, such as finite self-interactions, or other such deviations from the canonical LCDM paradigm.

\section{Summary and Conclusions}
\label{SecConc}

We have studied the tidal evolution of dark matter-dominated satellites embedded in LCDM dark matter subhalos as they orbit in the potential of a massive galaxy. Our study assumes that subhalos may be approximated by spherically symmetric, isotropic, cuspy Navarro-Frenk-White density profiles. We further assume that subhalo masses are much smaller than the host halo mass, so as to neglect the effects of dynamical friction.
As angular momentum plays only a subordinate role in determining if a dark matter particle gets stripped or remains bound (see Appendix~\ref{appendix:angmom}), we expect our results to hold not only for the isotropic models studied in this work, but also for mildly anisotropic models.
Our main findings may be summarized as follows.

\begin{itemize}
\item[(i)] Tides lead NFW subhalos to evolve following ``tidal tracks'' that describe the changes in characteristic size and velocity as tidal mass losses accumulate. These tidal tracks lead to a well defined asymptotic tidal remnant with characteristic crossing time set by the orbital time at pericentre, $T_\mathrm{mx}\approx T_\mathrm{peri}/4$. The asymptotic structure of the tidal remnant is well approximated by an exponentially truncated cusp.
  \item[(ii)]  Tidal evolution can be conveniently modelled in terms of binding energy; tides gradually truncate subhalos energetically, systematically removing particles with low binding energies. The bound remnant consists of particles initially more strongly bound than the energy threshold imposed by tides. 
 \item[(iii)]  For the majority of particles, the initial binding energy alone is sufficient to determine when and whether a particle will be stripped or not. Initial and final binding energies are strongly correlated, in a way that allows the structure of the remnant to be inferred as a function of tidal mass loss. These relations hold regardless of the orbital eccentricity of the subhalo.
 \item[(iv)] If gravitationally unimportant, the stellar components of tidally stripped NFW subhalos evolve according to how stars populate the initial energy distribution of the subhalo. The same gradual energy truncation applies to dark matter and stars, independent of the initial density structure or radial segregation of the stellar component. 
\item[(v)] The structure of stellar remnants retain memory of their initial structure, and, in particular, of how stars populate the most bound energy levels of the inner dark matter cusp. Their evolution is thus sensitively dependent on their initial structure and radial segregation.  Unlike the dark matter, stellar components do not follow unique ``tidal tracks'' in stellar mass, size, and velocity dispersion. 
\item[(vi)] The stellar component may be completely dispersed by tides if the stars don't populate the most bound energy levels of the dark matter cusp, making the potential existence of ``micro-galaxies'' critically dependent on how stars populate the most bound energy levels of an NFW subhalo. 
\item[(vii)] ``Tidally limited'' satellites, defined as those which have lost a substantial fraction of their initial stellar mass, have radii and velocity dispersions that trace directly the characteristic radius and velocity of the subhalo remnant. 
\end{itemize}

Most Local Group dwarfs have structural parameters consistent with them being embedded in cuspy NFW halos, but there are also a number whose dynamical masses  are below what is expected from LCDM. The properties of such systems, all of which are M31 or MW satellites, are usually assumed to result from the effects of tidal stripping. Our work sheds further insight into this interpretation.

Indeed, coupled with the assumption of a minimum halo mass, such as that suggested by hydrogen cooling limit considerations, our results place strong constraints on the size and velocity dispersion of tidally limited systems embedded in the remnants of cuspy NFW subhalos.  These constraints may be expressed as a firm lower limit on the velocity dispersion of an embedded stellar remnant of given size, or, equivalently, as an upper limit to the size of an embedded system of given velocity dispersion. Inspection of available data for dwarf galaxies in the Local Group, however, reveals a number of ultrafaint satellites that breach these limits.

The findings we report here imply that such systems cannot be understood as equilibrium systems embedded in the tidal remnants of cuspy NFW subhalos. This presents a serious challenge to our understanding of the formation and evolution of these unusual galaxies in LCDM. Reconciling systems like Crater~2, Antlia~2, or And 19 with LCDM seems to require substantial revision to at least one of the assumptions of this work. Either the reported observational parameters of such systems are in substantial error, or the systems are far from dynamical equilibrium, or they inhabit subhalos whose density profiles differ significantly from NFW.

Neither alternative is particularly attractive in the context of LCDM. It may indicate that baryons may affect the inner halo cusp even in extremely faint dwarfs or, more intriguingly, may indicate effects associated with the intimate nature of the dark matter, such as finite self-interactions or other such deviations from the canonical LCDM paradigm. Resolving this challenge will likely require a concerted numerical and observational effort designed to improve our understanding of the formation and evolution of some of the faintest galaxies known.

\section*{Acknowledgements}
RE acknowledges support provided by a CITA National Fellowship. Both RE and RI acknowledge funding from the European Research Council (ERC) under the European Unions Horizon 2020 research and innovation programme (grant agreement No. 834148). 
This work used the DiRAC@Durham facility managed by the Institute for Computational Cosmology on behalf of the STFC DiRAC HPC Facility (\url{www.dirac.ac.uk}). The equipment was funded by BEIS capital funding via STFC capital grants ST/K00042X/1, ST/P002293/1, ST/R002371/1 and ST/S002502/1, Durham University and STFC operations grant ST/R000832/1.

\section*{Data availability}
The data underlying this article will be shared on reasonable request to the corresponding author.

\footnotesize{
\bibliography{stars}
}

\appendix

\section{Stripping and angular momentum}
\label{appendix:angmom}

\begin{figure}
 \centering
     \includegraphics[width=4.25cm]{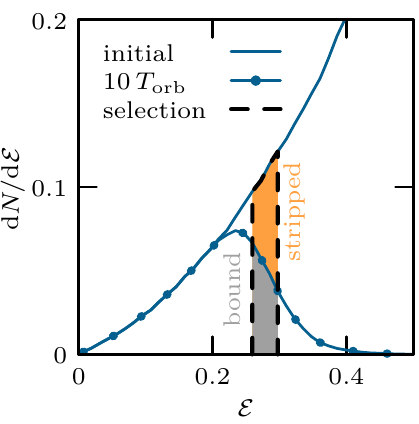}\includegraphics[width=4.25cm]{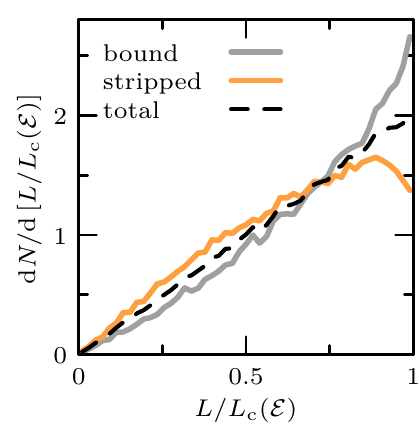} 
\caption{Left panel: Like Fig.~\ref{fig:E_DM_ICs}, in linear units, showing the initial energy distribution of subhalo model \emph{h-i}, and that of particles that form the bound remnant after 10 orbital periods (blue connected dots).
Half of the particles within the highlighted energy range (black dashed selection) get stripped (orange area), while the other half remains bound (grey area). 
Right panel: Angular momentum distribution $\diff N / \diff L$ of particles within the energy selection shown in the left panel. The angular momentum is expressed in units of the angular momentum of a circular orbit $L_\mathrm{c}$ of energy $\E$. The angular momentum distributions are normalised to integrate to $N=1$. The three distributions are remarkably similar. }
\label{fig:appendix:angmom}
\end{figure}

Tides gradually truncate NFW subhalos in energy, stripping the least-bound particles first, as shown in Fig.~\ref{fig:E_DM_ICs}. The truncation in energy is relatively sharp, and may be approximated using the `filter function'' of Eq.~\ref{eq:DM_filter}. This function gradually truncates the initial energy distribution beyond some energy $\E_\mathrm{mx,t}$. While in Fig.~\ref{fig:E_DM_ICs} almost all particles that fall ``to the left'' of the truncation energy $\E_\mathrm{mx,t}$ remain bound, there is a tail of particles towards less-bound energies where energy alone is not sufficient to tell whether these particles will be stripped, or remain bound. 
This is shown in the left panel of Fig.~\ref{fig:appendix:angmom}.

In the highlighted range of energies (black dashed selection), $0.25 \leq \E \leq 0.3$, half of particles get stripped, while the other half remains bound. The right panel of Fig.~\ref{fig:appendix:angmom} shows the (initial) angular momentum distributions of particles in the selected energy range, expressed in units of the angular momentum of a circular orbit $L_\mathrm{c}(\E)$ at energy $\E$.
The (initial) angular momentum distributions are remarkably similar. Only particles on nearly circular orbits exhibit a mild tendency to remain bound. Indeed, whereas $\sim 48\%$ of particles with $L/L_\mathrm{c} < 0.9$ remain bound (and $\sim 52\%$ are stripped), $\sim 61\%$ of  particles with $L/L_c  \geq 0.9$ remain bound (and $\sim39\%$ are stripped). This difference, however, affects fewer than $\sim 20\%$ of all particles in that energy bin.

While all models studied in this work are based on spherical, isotropic initial conditions that can be expressed as $f(\E)$ distribution functions, because of the small effect of angular momentum on determining if a particle gets stripped, we expect that our conclusions hold also for mildly anisotropic models (cast, e.g., as $f(\E,L_z)$ distribution functions).

\section{Orbital eccentricity}
\label{appendix:eccentricity}

\begin{figure}
 \centering
     \includegraphics[width=8.5cm]{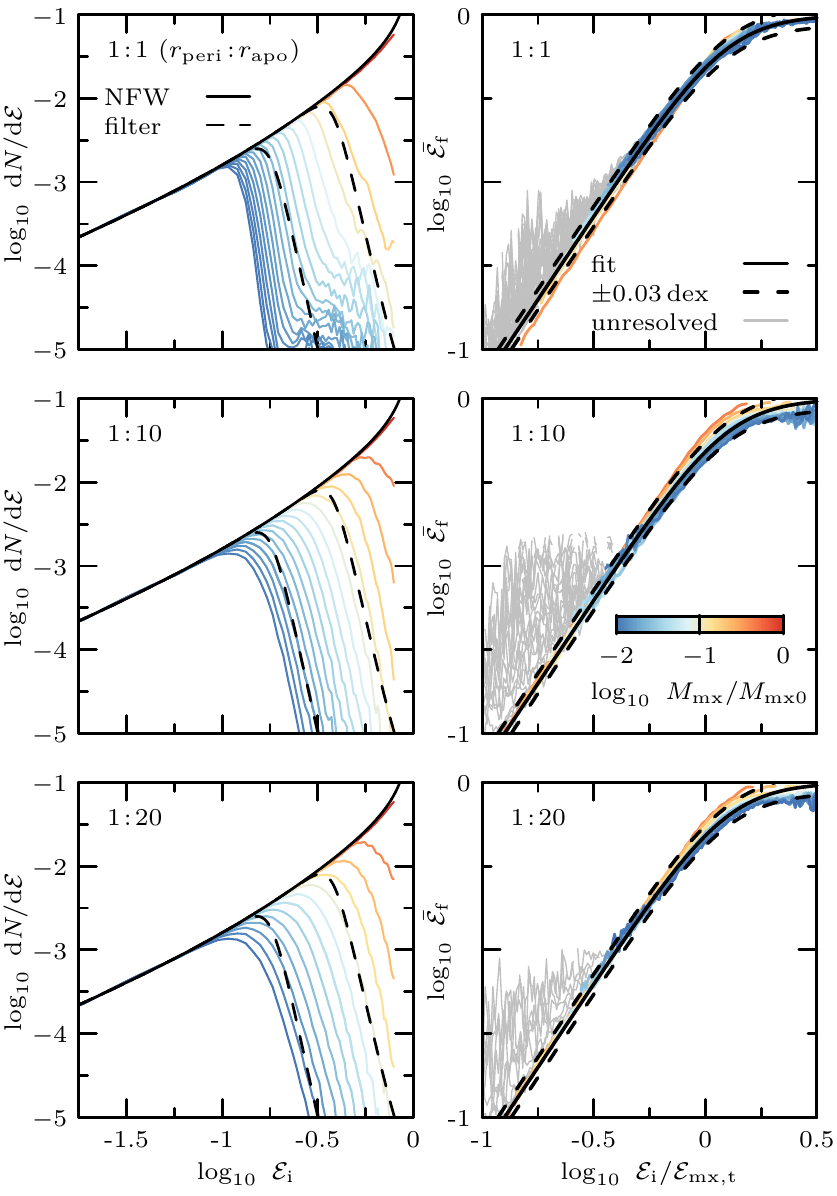}
\caption{Left column: Initial binding energy distribution of particles that remain bound to the subhalo after subsequent pericentric passages, like Fig.~\ref{fig:E_DM_ICs}. Right column: Median relation of the initial ($\E_\mathrm{i}$) to ``final'' ($\E_\mathrm{f}$) energy mapping, for initial energies, $\E_i/\E_\mathrm{mx,t}$, like Fig.~\ref{fig:energy_map_all}. The top row shows a model on a circular orbit, while the middle and bottom row show models on eccentric orbits with $r_\mathrm{peri}\rt r_\mathrm{apo}$ of $1\rt 10$ and $1 \rt$ 20, respectively. The initial structural properties of the different models are listed in the text. }
\label{fig:appendix:ecc}
\end{figure}

Section~\ref{sec:results_DM} discusses tidal stripping of NFW subhalos on an orbit with pericentre-to-apocentre ratio of $1\rt5$. This Appendix extends the analysis to circular ($1\rt1$) and more radial orbits ($1\rt10$, $1\rt20$). 
The circular orbit has a radius of $40\,\kpc$ and an orbital period of $\Torb = 1.12 \,\Gyrs$.
All eccentric orbits have an apocentre distance of $\rapo=200\,\kpc$, allowing the subhalo to reach equilibrium between subsequent pericentre passages. The $1\rt10$ and $1\rt20$ orbits consequently have pericentre distances of $\rperi = 20\,\kpc$ and $10\,\kpc$, and orbital periods of $2.31\,\Gyrs$ and $2.25\,\Gyrs$, respectively. 
Subhalo models were chosen from the \citet{EN21} set of simulations and, on the selected orbit, were stripped to less than one per cent of their initial mass. For the circular orbit, we used the same subhalo as for the $1\rt5$ of Section~\ref{sec:results_DM}, with $r\maxzero = 0.48\,\kpc$ and $V\maxzero = 3.0\,\kms$. For the $1\rt10$ and $1\rt20$ radial orbits, we used subhalos with $r\maxzero = 0.39\,\kpc$, $V\maxzero = 3.3\,\kms$ and $r\maxzero = 0.29\,\kpc$, $V\maxzero = 3.8\,\kms$. 

The left column of Fig. \ref{fig:appendix:ecc} shows the truncation in energy of subhalos on these three different orbits (as shown in Fig.~\ref{fig:E_DM_ICs} for a $1\rt5$ orbit). The filter function of Eq.~\ref{eq:DM_filter} (two examples are shown as black dashed curves in each panel) provides an approximative description for the shape of the energy truncation regardless of orbital eccentricity. The small deviations seen may contribute to the minor eccentricity dependence of tidal evolutionary tracks as shown in figure~6 of \citet{EN21}.

The right column of Fig. \ref{fig:appendix:ecc} shows that the initial-to-final energy mapping (as shown in Fig.~\ref{fig:energy_map_all} for a $1\rt5$ orbit) is well-described by Eq.~\ref{eq:energy_map}, regardless of eccentricity.

\section{Orbital periods}
\label{appendix:periods}

\begin{figure}
 \centering
     \includegraphics[width=8.5cm]{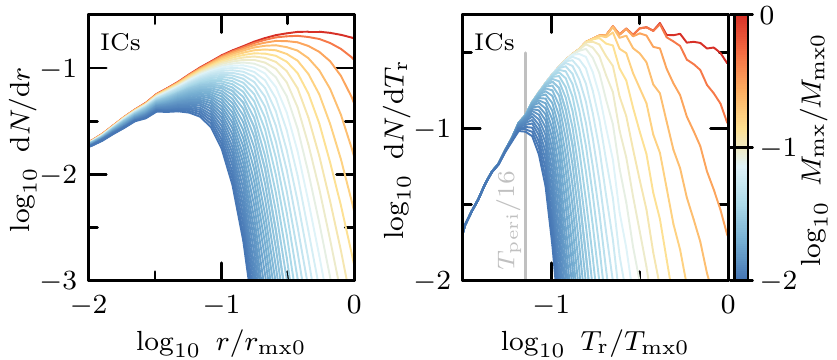} 
\caption{Radii $r/r\maxzero$ (left panel) and radial orbital periods $T_\mathrm{r}/T\maxzero$ (right panel) of particles in the initial conditions (model  \emph{h-i}) which remain bound at subsequent apocentre snapshots. Radii $r$ are normalised by the radius of maximum circular velocity $r\maxzero$, while periods $T_\mathrm{r}$ are normalised by the period $T\maxzero$ of a circular orbit at $r\maxzero$.
The remnant bound mass fraction $\Mmax/M\maxzero$ is colour-coded. Tides preferentially strip particles with larger orbital periods $T_\mathrm{r}$ and larger radii $r$ first.
The truncation in orbital periods imposed by tides is much steeper than the truncation in radius. Particles with initial orbital periods smaller than $\lesssim T_\mathrm{peri}/16$ form the asymptotic tidal remnant.
}
\label{fig:appendix:periods}
\end{figure}

As an NFW subhalo loses mass through tides, the least-bound particles get stripped first: the initial energy distribution $\diff N / \diff \E$ is sharply truncated beyond some energy $\E_\mathrm{mx,t}$ which depends on the remnant bound mass (see Fig.~\ref{fig:E_DM_ICs}). For isotropic, spherical NFW profiles, and for the majority of particles, energy is the only parameter determining whether it will be stripped, or not. 
Fig.~\ref{fig:appendix:periods} shows the distributions of initial radii $r$ (left panel) and radial orbital periods $T_r$ (right panel) of those particles which will form the bound remnant at subsequent apocentre snapshots. 
Orbital periods are computed from the initial energy $E$ and angular momentum $L$ of each $N$-body particle in the initial NFW potential \citep[see, e.g., ][eq. 3-16]{BT87}, 
\begin{equation}
\ T_r = 2 \int_{\rapo}^{\rperi} \diff r ~ \left\{  2 \left[E - \Phi_\mathrm{NFW}(r)\right] - L^2/r^2 \right\}^{-1/2} ~,
\end{equation}
and are normalized by the initial crossing time at the radius of maximum circular velocity, $T\maxzero = 2 \pi r\maxzero / \vmax$. 
The remnant mass fraction $\Mmax / M\maxzero$ is colour-coded. 

As $\Mmax / M\maxzero$ decreases, the bound remnant consists of particles which in the initial conditions were located at decreasing distance to the subhalo centre. Initial radii alone, however, do not determine whether a particle will remain bound or not. Indeed, at fixed radius in the left panel of Fig.~\ref{fig:appendix:periods}, some particles get stripped, but others remain bound. 

In contrast, the truncation in the initial radial orbital period $T_r$ imposed by tides is steep (see right panel of Fig.~\ref{fig:appendix:periods}), selecting most particles with radial orbital periods shorter than the truncation period. 
Tidal evolution stalls once the relaxed remnant has a crossing time of roughly one quarter of the host halo crossing time at pericentre,  $2 \pi \rmax / \vmax \approx \tperi/4$. In the initial conditions, this corresponds to a selection of particles with orbital periods shorter than $\approx \tperi/16$.

\section{Radius-dependent energy distribution}
\label{appendix:e_vs_r}

\begin{figure*}
 \centering
      \includegraphics[width=17.2cm]{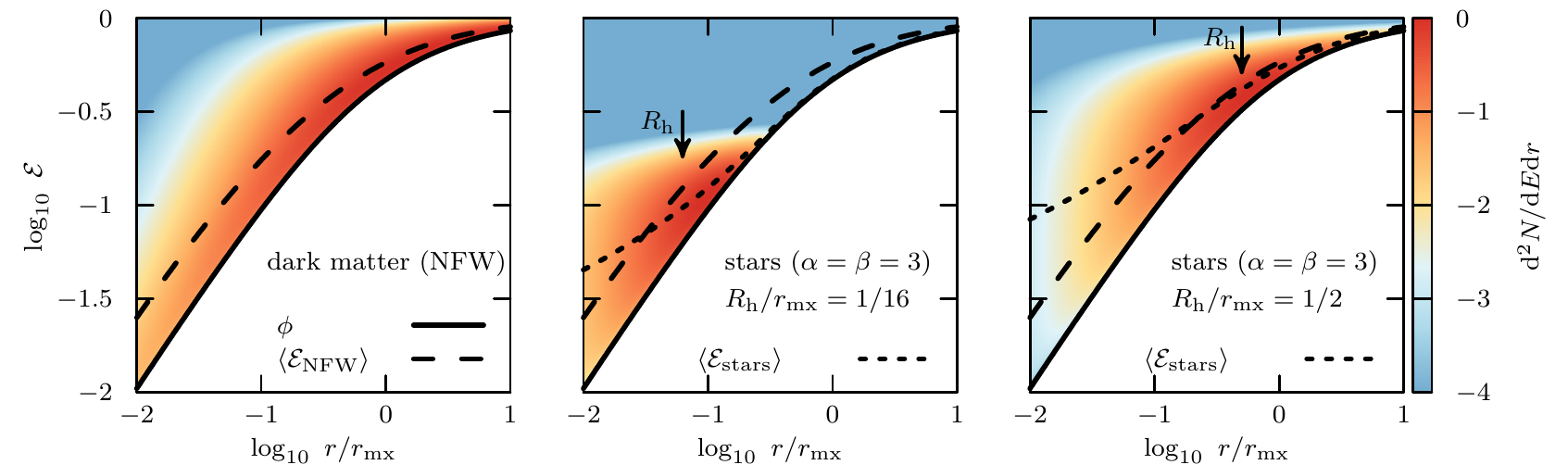} 
 \caption{Number of NFW dark matter (left panel) and stellar particles (middle and right panels) in the initial conditions as a function of radius $r$ and dimensionless energy $\E = 1 - E / \Phi_0$. The dimensionless potential $\phi \equiv 1-\Phi/\Phi_0$ is shown as a black solid line, while the average energies $\langle \E \rangle$ at fixed radius of dark matter and stellar particles are shown using dashed and dotted curves, respectively. The middle panel shows a deeply segregated stellar tracer ($\Rh / \rmax = 1/16$). There, at the half-light radius $\Rh$ (see arrow), stars are more bound than dark matter. Hence, if the system is exposed to a tidal field, initially, dark matter is stripped predominantly. In contrast, for the more extended stellar tracer ($\Rh / \rmax = 1/2$) shown in the right panel, at the half-light radius, stars and dark matter are similarly strongly bound, and both stars and dark matter may be stripped equally. }.
 \label{fig:e_dist_stars}
\end{figure*}

Depending on how deeply embedded a stellar tracer is within its dark matter halo, the peak of its energy distribution shifts. The deeper embedded a stellar tracer is, the more its peak energy shifts towards strongly-bound energies. This has qualitative consequences for the tidal evolution of stellar structural parameters, which are discussed in this appendix. Figure \ref{fig:e_dist_stars} shows, in the initial conditions, the number of 
dark matter particles (left panel) and stellar particles (middle and right panel for tracers with $R_\mathrm{h0}/r\maxzero=1/16$ and $1/2$, respectively) for a given radius $r$ and energy $\E = 1-E/\Phi_0$. 
The probability $P(r)$ of a dark matter particle to be located at a radius $r$ follows directly from the NFW density profile, 
\begin{equation}
 P(r) = 4 \pi r^2 \rho_\mathrm{NFW}(r) ~,
\end{equation}
while the probability for a particle to have an energy $E$ at a fixed radius $r$ equals
\begin{equation}
 P(E|r) = \left\{ 2 \left[ E - \Phi(r) \right] \right\}^{1/2} r^2 f(E)~,
\end{equation}
where $f(E)$ is the NFW distribution function, obtained through Eddington inversion \citep[see, e.g.,][]{EP20}.
The probability for a dark matter particle to be located at a radius $r$ and to have an energy $E$ is then $P(E,r) = P(r) P(E|r)$. For stellar particles, this probability needs to be multiplied by the stellar tagging probability of Eq.~\ref{eq:stellartag}. 

No particles located at a radius $r$ may have an energy $\E$ lower than the potential $\phi(r) = 1 - \Phi(r)/\Phi_0$ at that radius, shown as a solid black line.
The dark matter (NFW) particles span a wide range in radii and energies, with substantial mass located at high binding energies within the central dark matter density cusp. In contrast, the initial energy distribution of stellar tracers is more localized, i.e. the range of radii and energies where most particles are located is narrower. 

The average energy at a fixed radius for dark matter particles is shown as a dashed curve in all panels, while the average energy of each stellar tracer is shown using a dotted line. For the more deeply embedded stellar system, at the half-light radius $\Rh$ (corresponding approximatively to the radius where most particles are located, see colour coding), stellar particles are on average more strongly bound than dark matter. Hence tidal stripping will initially preferentially remove dark matter. This causes the mass enclosed within the stellar half-light radius to drop (and thereby the stellar velocity dispersion to decrease), and the half-light radius to expand during relaxation. On the other hand, for the more extended stellar tracer, at the half-light radius, stars and dark matter have similar binding energies. As tides strip the system, both dark matter and stars are hence lost from the beginning, and $\Rh$ drops, consistent with the models shown in Fig.~\ref{fig:tracks_mmx}.

\section{Isotropic stellar models}
\label{appendix:isotropic_models}
\begin{figure}
 \centering
   \includegraphics[width=8.5cm]{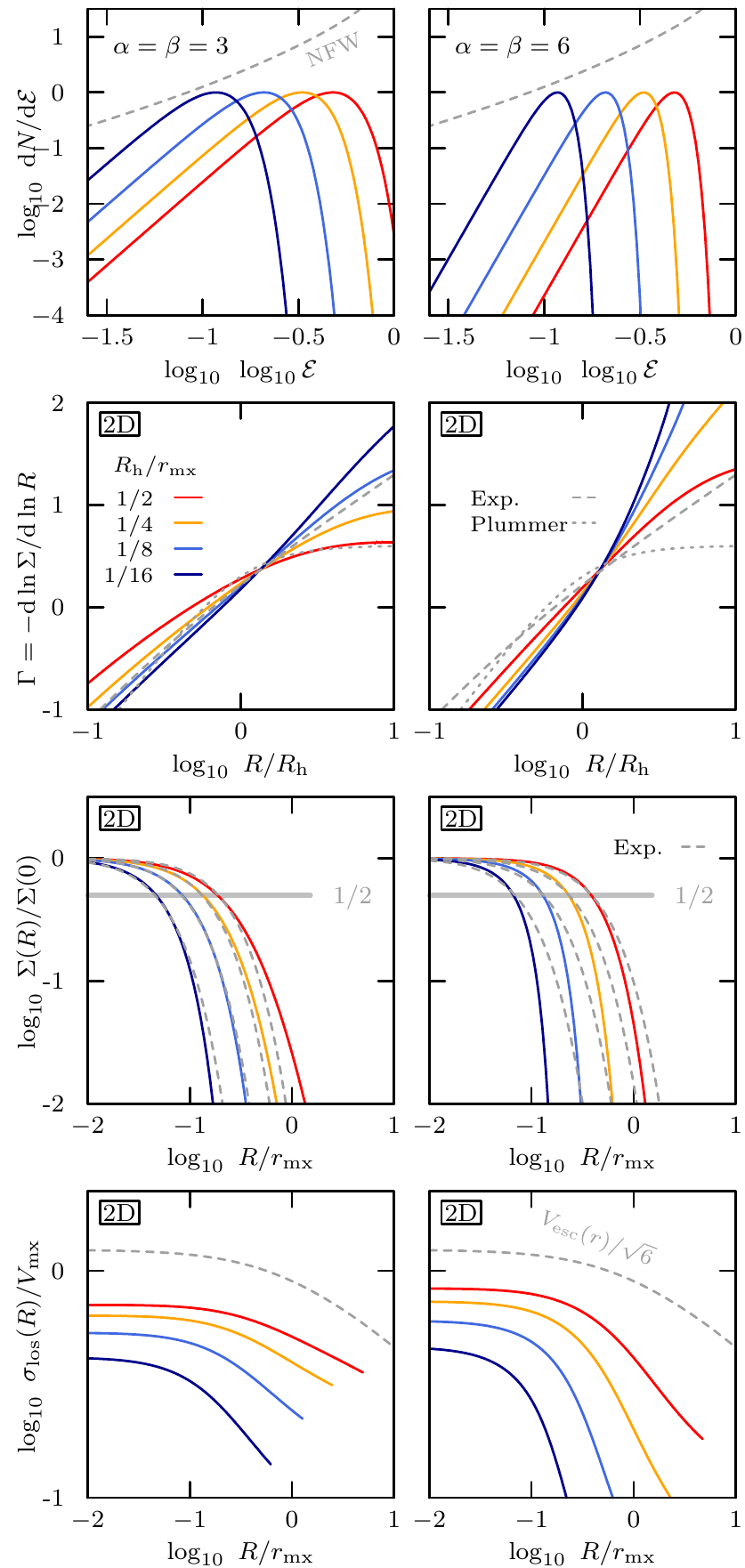} 
\caption{Exploration of the structure and kinematics of spherical, isotropic stellar models that follow the energy distribution of Eq.~\ref{eq:this_dnde} (with $\alpha=\beta=3$ in the left column, and $\alpha=\beta=6$ in the right column), embedded in an NFW dark matter halo.
The energy distributions of stellar tracers with half-light radii of $\Rh/\rmax = 1/2$ (red), $1/4$ (orange), $1/8$ (light blue), $1/16$ (dark blue), where $\rmax$ is the radius of maximum circular velocity of the underlying dark matter halo, are shown in the top panel. 
The logarithmic slope $\Gamma = \diff \ln \Sigma / \diff \ln R$ of the surface brightness profiles is compared against that of Plummer and exponential models (grey dashed curves), showing that the $\alpha=\beta=3$ model approximatively resembles an exponential profile for radii smaller than the 2D half-light radius $\Rh$ (2nd panel from top). The corresponding surface brightness profiles are compared against exponential profiles of the same core radius (3rd panel from top). The line-of-sight velocity dispersion profiles are shown in the bottom panel in units of the maximum circular velocity $\vmax$ of the underlying dark matter halo.}
\label{fig:appendix:isotropic_sol}
\end{figure}

This appendix presents the structural properties of spherical stellar systems with isotropic velocity dispersion, embedded in NFW dark matter halos, with energy distribution $\diff N_\star  / \diff \E$ given by Eq.~\ref{eq:this_dnde}. Figure \ref{fig:appendix:isotropic_sol} shows the energy distributions (top panel), logarithmic slopes of the surface brightness profiles (2nd row), surface brightness profiles (3rd row) and line-of-sight velocity dispersion profiles (bottom row) for models with $\alpha=\beta=3$ (left column), and $\alpha=\beta=6$ (right column). Profiles for four different segregations, $\Rh/\rmax = 1/2$ (red), $1/4$ (orange), $1/8$ (light blue) and $1/16$ (dark blue), are shown. 

Structural properties are computed directly from the distribution function $f(\E) \propto  (\diff N_\star  / \diff \E)/ p(\E) ~  $, where $p(\E)$ is the phase-space volume accessible in an NFW potential to a particle with energy $\E$ (i.e., the density of states). 

The more deeply embedded the stellar profile is within its dark matter halo, the further the maximum of the energy distribution is shifted towards high binding energies. 
The range of energies selected by the $\alpha=\beta=3$ model is much wider than that of the $\alpha=\beta=6$ model. For equal values of $\Rh/\rmax$, the energy $\E_\star$ at which the distribution peaks is near identical between the two models. 

The exact shape of the stellar density profile depends on where it is embedded inside the dark matter halo. $\alpha=\beta=3$ models approximately resemble exponential density profiles, while $\alpha=\beta=6$ profiles have larger core radius (i.e., a radius $R_\mathrm{c}$ so that $\Sigma(R_\mathrm{c}) = \Sigma(0)/2$) at equal half-light radius. 

The more deeply embedded a stellar tracer is within its dark matter halo, the lower is its central velocity dispersion relative to the subhalo characteristic velocity $\vmax$ (bottom panel). A constant density tracer of infinite extent has a 3D velocity dispersion profile equal to $V_\mathrm{esc}(r)/\sqrt{2}$, hence, 
for reference, the escape velocity profile of an NFW subhalo is shown in the bottom panel of Fig.~\ref{fig:appendix:isotropic_sol} (re-scaled by a factor of $\sqrt{3}$ to compare against the 1D line-of-sight velocity profiles shown).

\section{Tidal tracks}
\label{appendix:tidal_tracks}
\begin{figure}
 \centering
     \includegraphics[width=8.5cm]{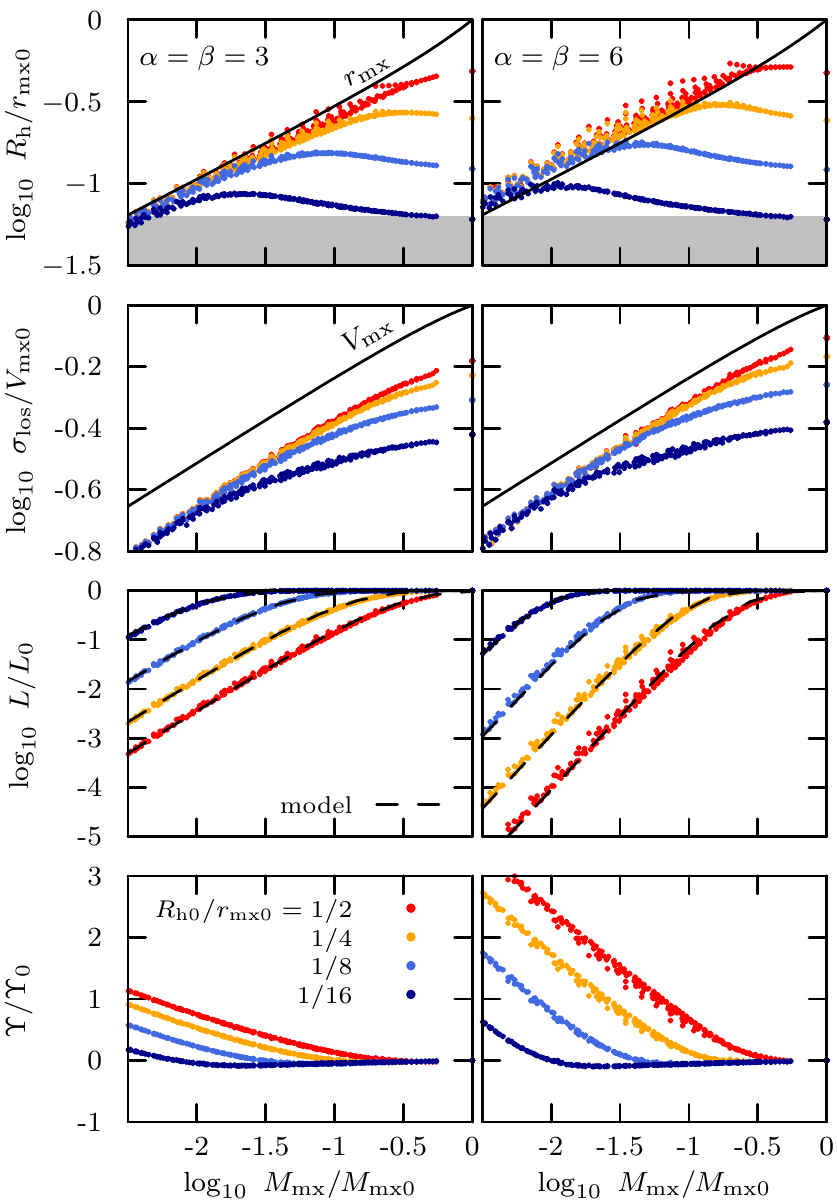} 
\caption{Evolution of stellar structural parameters as a function of remnant dark matter bound mass fraction $\Mmax/M\maxzero$ for stellar tracers with initial structure as shown in Fig.~\ref{fig:appendix:isotropic_sol}, embedded in all subhalos of the \citet{EN21} set of simulations on orbits with pericentre-to-apocentre ratio of $1\rt5$ (for details, see Section~\ref{sec:nummethods}). Different initial segregations are shown using different colours. As tides strip the dark matter halo, the stellar half-light radii $\Rh$ asymptotically approach the dark matter characteristic radius $\rmax$ (black solid curve, top panel), and their line-of-sight velocity dispersions $\sigmalos$ evolve in parallel to the dark matter characteristic velocity $\vmax$ (black solid curve, 2nd panel from top). The luminosity $L$ drops, and the rate of stellar mass loss depends significantly on the inner slope $\alpha$ of the underlying energy distribution (3rd panel from top). As $L$ drops, the mass-to-light ratio $\Upsilon$ (averaged here within the 2D half-light radius) increases (bottom panel).}
\label{fig:tracks_mmx}
\end{figure}

Structural parameters of dwarf spheroidal galaxies evolve differently under tidal stripping depending on how deeply embedded they are within the dark matter halo, and which energies they populate within the halo. Fig.~\ref{fig:tracks_mmx} shows the evolution of half-light radius $\Rh$ (top panel), line-of-sight velocity dispersion $\sigmalos$ (2nd panel from top), luminosity $L$ (3rd panel from top) and mass-to-light ratio averaged within the half-light radius $\Upsilon = L/[2 M(<\Rh)]$, as a function of remnant dark matter mass fraction $\Mmax/M\maxzero$. The evolution is shown for stellar tracers with initial half-light radii of $\Rhzero/r\maxzero = 1/2$ (red), $1/4$ (orange), $1/8$ (light blue), $1/16$ (dark blue). Each point corresponds to one apocentre snapshot, taken from the set of simulations described in Sec.~\ref{sec:nummethods}. The evolution of the subhalo characteristic radius $\rmax$ and velocity $\vmax$ are shown using solid black curves. Stellar tracers with $\alpha=\beta=3$ and $\alpha=\beta=6$ (Eq.~\ref{eq:this_dnde}) energy distribution are shown in the left column and right column, respectively. The full initial profiles for these tracers are shown in Fig.~\ref{fig:appendix:isotropic_sol}.

In the ``tidally limited'' regime, the stellar half-light radii $\Rh$  trace the dark matter characteristic size $\rmax$ (for the $\alpha=\beta=3$ and $\alpha=\beta=6$ models, we find in the tidally limited regime $\Rh \approx 0.93\,\rmax$ and  $\Rh \approx 1.26\,\rmax$, respectively). Similarly, the stellar velocity dispersion $\sigmalos$ evolves parallel to the dark matter characteristic velocity $\vmax$ (for $\alpha=\beta=3$ and $\alpha=\beta=6$, we find $\sigmalos \approx 0.70\,\vmax$ and  $\sigmalos \approx 0.76\,\vmax$). Once a stellar tracer has been stripped to the size of $\Rh \approx \rmax$, it becomes impossible to reconstruct its original extent from structural properties alone.

The rate at which the stellar luminosity $L$ decreases depends crucially on the energy distribution of the stars within the dark matter halo: luminosity drops faster for stellar tracers with energy distributions that have a steeper slope towards the most-bound energies (i.e., larger $\alpha$ in Eq.~\ref{eq:this_dnde}). The total luminosity of the remnant is given by all stars with energies more tightly bound than the tidal truncation. As the same ``filter function'' (Eq.~\ref{eq:DM_filter}) can be applied to both dark matter and stars to select those particles which remain bound for a given energy truncation $\E_\mathrm{mx,t}$, the remnant luminosity may be computed through the integral
\begin{equation}
 L = \int_0^1 \diff \E ~   {\left. \diff N_\star / \diff \E \right|_\mathrm{i}  \over  1 + \left(a\, \E / \E_\mathrm{mx,t} \right)^{b}},
\end{equation}
with $a\approx0.85$, $b\approx 12$ (as in Eq.~\ref{eq:DM_filter}). In the above equation, $ \diff N_\star  / \diff \E |_\mathrm{i}$ denotes the initial stellar energy distribution (normalised to give, when integrated, the initial luminosity), and $\E_\mathrm{mx,t}$ may be computed from the remnant bound mass fraction $\Mmax/M\maxzero$ through Eq.~\ref{eq:Mmx_to_E}. The result of this calculation is shown as a dashed curve (``model'') in the third row of Fig.~\ref{fig:tracks_mmx}.

For order-of-magnitude estimates, a simple power-law approximation to this integral may be used for heavily-stripped systems. For NFW (Eq.~\ref{eq:NFW}) density profiles and those of an exponentially truncated cusp (Eq.~\ref{eq:TruncCusp}), the energy distribution $\diff N_\mathrm{DM}/ \diff \E  $ is well-approximated by a power-law of slope $\approx1$ towards the most-bound energies, i.e. $\diff N_\mathrm{DM}/ \diff \E \propto \E$ for $\E \ll 1$. Assuming that the truncation happens sharpy at an energy $\E_\mathrm{mx,t} \ll 1 $, and that the density profile has converged to that of an exponentially truncated cusp, this gives $\Mmax \propto M \approx \int_0^{\E_\mathrm{mx,t}} \diff \E ~ \diff N_\mathrm{DM}/\diff \E  \propto \E_\mathrm{mx,t}^2$. Similarly, for stellar tracers with an energy distribution parametrised through Eq.~\ref{eq:this_dnde} and $\E_\mathrm{mx,t} \ll \E_\star$,  $L \approx \int_0^{\E_\mathrm{mx,t}} \diff \E ~ \E^\alpha \propto {\E_\mathrm{mx,t}}^{\alpha +1}$. Consequently $L \propto M^{(\alpha+1)/2}$ for highly stripped stellar systems.

\section{Empirical model for evolved stellar parameters}
\label{appendix:step-by-step}
In this appendix, we discuss how to construct the relaxed energy distribution of a tidally stripped (stellar) tracer in a cold dark matter subhalo.
For this, we make use of the observation that stellar energy distributions are subject to the same tidal truncation as the underlying NFW dark matter halo (Sec.~\ref{sec:results_stars}), and of the energy mapping between the initial NFW halo and the and the stripped, relaxed system discussed in Sec.~\ref{sec:energy_map}. 

Given an initial stellar energy distribution, $\diff N_\star / \diff \E |_\mathrm{i}$, as well as the desired remnant mass fraction of the underlying NFW dark mater halo, $\Mmax/M\maxzero$, the procedure involves the following steps:

\begin{itemize}
\item[\bf{(i)}] For the desired remnant bound mass fraction $\Mmax/M\maxzero$ of the underlying dark matter subhalo, the tidal truncation energy $\E_\mathrm{mx,t}$  is computed from Eq.~\ref{eq:Mmx_to_E}.
\end{itemize}
\begin{itemize}
\item[\bf{(ii)}] The energy distribution $\diff N_\star / \diff \E |_\mathrm{i,t}$ of those particles in the initial conditions which remain bound at the given tidal truncation energy $\E_\mathrm{mx,t}$ follows from applying the filter of Eq.~\ref{eq:DM_filter} to the initial energy distribution.
\end{itemize}
\begin{itemize}
\item[\bf{(iii)}] Finally, we need to model the relaxation process to pass from the initial truncated energy distribution $\diff N_\star / \diff \E |_\mathrm{i,t}$ to the relaxed, equilibrium system $\diff N / \diff \E |_\mathrm{f}$. 
To take into account the scatter\footnote{Ignoring any scatter around the average energy mapping of Eq.~\ref{eq:energy_map}, the final energy distribution of the relaxed system would now follow simply from applying the average energy mapping to the truncated distribution in the initial conditions, i.e. $ \left. {\diff N_\star}/{\diff \E} \right|_\mathrm{f} =  \left. \diff N_\star\left( \bar \E_\mathrm{f}^{-1}(\E_\mathrm{f}) \right) / \diff \E \right|_\mathrm{i,t}~ | \diff \bar \E_\mathrm{f}^{-1} / \diff \E_\mathrm{f}|$. Here, $\bar \E_\mathrm{f}^{-1}(\E_\mathrm{f})$ is the inverse of Eq.~\ref{eq:energy_map}, and $|\diff \E_\mathrm{f}^{-1} / \diff \E_\mathrm{f}|$ is the absolute value of the corresponding Jacobian.}
in the energy mapping shown in the top panel of Fig.~\ref{fig:energy_map}, we compute the final energy distribution through the convolution 
\begin{equation}
  \left. \frac{\diff N_\star}{\diff \E} \right|_\mathrm{f} = \int_{0}^{1} \diff \E_\mathrm{i} ~ \left. \frac{\diff N_\star}{\diff \E}   \right|_\mathrm{i,t} ~ \mathrm{Lognormal}\left( \bar \E_\mathrm{f}(\E_\mathrm{i}) , \mathrm{0.03\,dex} \right)   ~.
\end{equation}
\end{itemize}
where $ \mathrm{Lognormal}\left(\bar \E_\mathrm{f}(\E_\mathrm{i}) , \mathrm{0.03\,dex} \right)  $ is a base-10 log-normal of width 0.03 dex modelling the scatter around the average relation of Eq.~\ref{eq:energy_map}.

An implementation of this procedure is made available online\footnote{\url{https://github.com/rerrani/tipy}}, taking as input the (stellar) initial energy distribution (parametrised through Eq.~\ref{eq:this_dnde}) as well as a desired dark matter remnant mass fraction $\Mmax/M\maxzero$, and returning the evolved stellar energy distribution.

\label{lastpage}

\end{document}